\documentclass{elsart}
\def\Lapar{\triangle_\|}
\def\napar{\nabla_\|}

\def\naperp{\nabla_\perp}
\def\neight{\frac{n{+}8}{9}}
\def\ntwo{\frac{n{+}2}{3}}
\def\nfive{\frac{n{+}5}{6}}

\def\hgf{\mathop{_{1\!}{F}_{2}}\nolimits}
\def\hgpfq{\mathop{_{p}{F}_{q\!}}\nolimits}
\def\hgtwofthree{\mathop{_{2}{F}_{3\!}}\nolimits}
\def\hgf{\mathop{_{1\!}{F}_{2\!}}\nolimits}
\def\ppsiq{\mathop{_{p\!}\Psi_{q}}\nolimits}
\def\modStruveL{\mathop{\mathbf{L}}\nolimits}
\def\res{\mathop{\mathrm{Res}}\nolimits}
\usepackage{amssymb,amsbsy,graphicx,texdraw}
\journal{Nuclear Physics B}
\begin{document}
\begin{frontmatter}
\title{Two-loop renormalization-group analysis of critical behavior at  
$\boldsymbol{m}$-axial Lifshitz points
}
\author{M.\ Shpot} \and
\author{H. W. Diehl}
\address{Fachbereich Physik, Universit{\"a}t
Essen, D-45117 Essen, Federal Republic of Germany}
\begin{abstract}
We investigate the critical behavior that
$d$-dimensional systems with short-range forces and
a $n$-component order parameter exhibit at Lifshitz points
whose wave-vector instability occurs
in an $m$-dimensional isotropic subspace of ${\mathbb R}^d$.
Utilizing dimensional regularization and minimal subtraction of poles
in $d=4+{m\over 2}-\epsilon$ dimensions,
we carry out a two-loop renormalization-group (RG) analysis
of the field-theory models
representing the corresponding universality classes.
This gives the beta function $\beta_u(u)$
to third order,
and the required renormalization factors as well as the
associated RG exponent functions to second order, in $u$.
The coefficients of these series are reduced to $m$-dependent
expressions involving single integrals,
which for general (not necessarily integer) values of $m\in (0,8)$
can be computed numerically,
and for special values of $m$  analytically.
The  $\epsilon$ expansions of the critical exponents
$\eta_{l2}$, $\eta_{l4}$, $\nu_{l2}$,  $\nu_{l4}$,
the wave-vector exponent $\beta_q$, and the correction-to-scaling
exponent are obtained to order $\epsilon^2$.
These are used to estimate their values for $d=3$.
The obtained series expansions are shown to
encompass both isotropic limits $m=0$ and $m=d$.
\end{abstract}
\end{frontmatter}

\section{Introduction}

The modern theory of critical phenomena \cite{Fis74,DG76,Fis83}
has taught us that the
standard $|\boldsymbol{\phi}|^4$ models with a
$n$-component order-parameter field
$\boldsymbol{\phi}=(\phi_i,\;i=1,\ldots,n)$
and $O(n)$ symmetric action
have significance which extends far beyond the
models themselves: They describe the long-distance physics of
whole classes of  microscopically distinct systems
near their critical points.
In fact, they are the simplest continuum models representing
the $O(n)$ universality classes of $d$-dimensional systems with
short-range interactions whose dimensions $d$ exceed the
lower critical dimension $d_*$ ($=1$ or $2$) for the appearance
of a transition to a phase with long-range order, and are
less or equal than the upper critical dimension $d^*=4$
(above which Landau theory  yields the correct asymptotic critical behavior).
By investigating these models via sophisticated field theoretical
methods \cite{ZJ96}, impressively accurate results have been obtained
for universal quantities such as critical exponents and universal amplitude ratios.

A well-known crucial feature of these models is their scale (and conformal)
invariance at criticality: The order-parameter density
behaves under scale transformations asymptotically as
\begin{equation}
\boldsymbol{\phi}(\ell\boldsymbol{x})\sim\ell^{-x_\phi}\,
\boldsymbol{\phi}(\boldsymbol{x})
\end{equation}\label{scinv}
 in the infrared limit $\ell\to 0$,
where $x_\phi=(d-2+\eta)/2$ is the scaling dimension of $\boldsymbol{\phi}$.
This scale invariance is {\it isotropic\/}
inasmuch as all $d$ coordinates of the position
vector $\boldsymbol{x}$ are rescaled in the same fashion.

There exists, however, a wealth of phenomena that exhibit scale invariance
of a more general, \emph{anisotropic}  nature. Roughly speaking,
one can identify four different categories:
(i) static critical behavior in anisotropic equilibrium systems
such as dipolar-coupled uniaxial ferromagnets \cite{Aha76} or systems with  
Lifshitz points \cite{Hor80,Sel92},
(ii) anisotropic critical behavior in stationary states of nonequilibrium systems
(like those of driven diffusive systems \cite{SZ95} or encountered in
stochastic surface growth \cite{Kru97}),
(iii) dynamic critical phenomena of  systems near thermal equilibrium
\cite{HH77},
and (iv) dynamic critical phenomena in nonequilibrium systems \cite{SZ95}.

In the cases of  the first two categories, the coordinates $\boldsymbol{x}$
can be divided into  two (or more) groups that  scale in a different fashion.
Writing $\boldsymbol{x}=({\boldsymbol{x}}_\|,{\boldsymbol{x}}_\perp)$, we call
these parallel and perpendicular, respectively. Instead of Eq.~(\ref{scinv})
one then has
\begin{equation}\label{asci}
\boldsymbol{\phi}(\ell^\theta {\boldsymbol{x}}_\|,
\ell{\boldsymbol{x}}_\perp)\sim\ell^{-x_\phi}\,
\boldsymbol{\phi}({\boldsymbol{x}}_\|,
 {\boldsymbol{x}}_\perp)\;,
\end{equation}
where $\theta$, the anisotropy exponent, differs from one. Categories
(iii) and (iv) involve genuine time-dependent phenomena for which time
typically scales with a nontrivial power of the length rescaling factor $\ell$.
For phenomena of category  (ii), one cannot normally avoid to deal
also with the time evolution.
This is because
a fluctuation-dissipation theorem generically does  \emph{not} hold
for such nonequilibrium systems;  their stationary-state distributions
are not  fixed by given Hamiltonians of equilibrium systems and hence
have to be determined from the long-time limit of their time-dependent
distributions in general.

Category (i) provides very basic examples of systems
exhibiting anisotropic scale invariance whose advantage is that
they can be investigated entirely within the framework of equilibrium
statistical mechanics. The particular example we shall be concerned with
in this paper is the familiar continuum model for an $m$-axial Lifshitz point,
defined by the Hamiltonian
\begin{equation}\label{Ham}
{\mathcal{H}}={\int}\!{d^d}x
{\left\{
{\rho_0\over 2}\,{({{\nabla}_\parallel}
 \boldsymbol{\phi})}^2
+{\sigma_0\over 2}\,{(\Lapar
 \boldsymbol{\phi})}^2
+\frac{1}{2}\,{({{\nabla}_\perp}
 \boldsymbol{\phi})}^2
+{\tau_0\over 2}\,
\boldsymbol{\phi}^2+\frac{u_0}{4!}\,|\boldsymbol{\phi}|^4\right\}}\;.
\end{equation}
Here $\boldsymbol{\phi}(\boldsymbol{x})=(\phi_i(\boldsymbol{x}))_{i=1}^n$
is an $n$-component order-parameter field where
$\boldsymbol{x}=(\boldsymbol{x}_\|,\boldsymbol{x}_\perp)\in{\mathbb{R}}^{m}
\times{\mathbb{R}}^{d-m}$. The operators $\napar$, $\naperp$, and $\Lapar$
 denote the
$d_\|=m $ and $d_\perp=d{-}m$ dimensional parallel and perpendicular components
of the gradient operator $\nabla$ and the associated Laplacian $\Lapar={\napar}^2$,
respectively.
The parameters $\sigma_0$
and $u_0$ are assumed to be positive. At zero-loop order (Landau theory),
the Lifshitz point is located at $\rho_0=\tau_0=0$.

We recall that a  Lifshitz point is a critical point where a disordered phase,  
a spatially
uniform ordered phase, and a spatially modulated ordered phase meet.
For further background and  extensive lists of references, the reader is
referred to review articles by Hornreich  \cite{Hor80} and Selke \cite{Sel92}
and to a number of more recent papers  
\cite{NF95,NAF97,MC98,MC99,BF99,FKB00,Lei00,DS00b}.

An attractive feature of the model (\ref{Ham}) is that the parameter $m$
can be varied. It was studied many years ago \cite{HLS75b,Muk77,HB78,SG78}
by means of an $\epsilon$ expansion about the
upper critical dimension
\begin{equation}\label{ucd}
d^*(m)=4+{m/2}\;,\quad m\le 8\,.
\end{equation}
The order-$\epsilon$ results for the
correlation-length exponents $\nu_{l 2}$ and
$\nu_{l 4}$, first derived by Hornreich et al.~\cite{HLS75b}, are generally accepted.
Yet long-standing controversies existed on the $\epsilon^2$ terms of the
correlation exponents $\eta_{l 2}$ and $\eta_{l 4}$ and the wave-vector
exponent $\beta_q$: Mukamel \cite{Muk77} gave results
for all $m$ with $1\le m\le 8$. These agreed with what
Hornreich and Bruce \cite{HB78} found in the uniaxial
case $m=1$  via an independent calculation, but were at variance
with Sak and Grest's \cite{SG78} for $m=2$ and $m=6$ (who investigated only these
special cases).

More recently, Mergulh{\~a}o and Carneiro \cite{MC98,MC99} presented a
reanalysis of the problem based on renormalized field theory and dimensional
regularization. Treating explicitly only the cases $m=2$ and $m=6$, they
recovered Sak and Grest's results for $\eta_{l 2}$ and $\eta_{l 4}$, but did not
compute $\beta_q$.  They analytically continued in $d_\|$
rather than in $d_\perp$ and, fixing the latter at
$d_\perp=4-{m/2}$ while taking the former as
 $d_\|=m-\epsilon_\|$, with $m=2$ or $6$, they also derived
the expansions of the correlation-length exponents
$\nu_{l 2}$ and $\nu_{l 4}$ to order $\epsilon_\|^2$.

The purpose of the present paper is to give
a full two-loop renormalization group (RG) analysis of the model (\ref{Ham})
for  general, not necessarily integer values of $m\in(0,8)$ in   
$d=d^*(m)-\epsilon$ dimensions.\nocite{dAL01}%
\footnote{Another two-loop calculation was recently attempted by de Albuquerque
and Leite \cite{dAL01,dAL01a}. In their evaluation of  two-loop graphs---e.g.,  
of  the
last graph of $\Gamma^{(4)}$ shown in Eq.~(\ref{Gamma4}),--- they replaced the  
integrand of the double
momentum integral by its value on a line.
We fail to see why such  a procedure, by which more or less
arbitrary numbers can be produced, should give meaningful results.
Let us also emphasize that  using such `approximations'  leads to the
following problem: Unless the corresponding  `approximations' are made for
higher-loop graphs involving this two-loop graph as a subgraph, pole terms
that \emph{cannot be absorbed by local counterterms} are expected to remain
because  they will not be canceled automatically through the subtractions provided
by counterterms of  lower order.
}
As a result we obtain the $\epsilon$ expansions of all
critical exponents $\eta_{l 2}$, $\eta_{l 4}$, $\beta_q$, $\nu_{l 2}$, and
$\nu_{l 4}$ to order $\epsilon^2$. In a previous paper \cite{DS00b},
hereafter referred to as I, we
have shown how to overcome the severe technical difficulties that had
hindered analytical progress in this field and prevented a resolution of the
above-mentioned controversy for so long. Working directly in position space
and exploiting the scale invariance of the free propagator at the Lifshitz point,
we were able to compute the two-loop graphs of the two-point vertex function
$\Gamma^{(2)}$ and $\Gamma^{(2)}_{(\nabla_\|\phi)^2}$,
its analog with an insertion of ${\int}\!\d^dx\,{(\nabla_\|\phi)^2/2}$.
Together with one-loop results, these suffice for determining
the exponents $\eta_{l 2}$, $\eta_{l 4}$, and $\beta_q$
to order $\epsilon^2$. In order to obtain the correlation-length exponents
$\nu_{l2}$ and $\nu_{l4}$ to this order in $\epsilon$ we must
compute the two-loop graphs of the four-point vertex function
$\Gamma^{(4)}$ and of $\Gamma^{(2)}_{\phi^2}$.

Our results are of importance to recent work on the generalization
of conformal invariance to anisotropic scale invariant systems  
\cite{Hen97,Hen99,PH01}.
Some time ago Henkel \cite{Hen97} proposed a new set of infinitesimal transformations
generalizing scale invariance for systems of this kind with an anisotropy exponent
$\theta =2/\wp$, $\wp=1,2,\ldots$. He pointed out that the case $\wp=4$,  
$\theta=1/2$,
is realized for the Lifshitz point of a spherical  ($n\to\infty$) analog of   
the ANNNI model \cite{HLS75,FH93}, and that the same $m$-independent value of  
$\theta$
in Ref.~ \cite{HLS75}  was found
to persist to first order in $\epsilon$  for the Lifshitz point of the model  
(\ref{Ham}).
However, as can be seen from Eq.~(\ref{thetaeps}) below [and Eq.~(84) of I],
$\theta$ \emph{deviates from\/} $1/2$ at order $\epsilon^2$.
This shows that the Lie algebra discussed in Ref.~\cite{Hen97} cannot
strictly apply below the upper critical dimension ($\epsilon>0$) if $n$ is finite,
except in  the trivial Gaussian case $u_0=0$.

The remainder of this paper is organized as follows.
In the next section we recapitulate the scaling form of  the
free propagator for $\tau_0=\rho_0=0$. We give the explicit form
of its scaling function as well as those of  similar quantities, and
discuss their asymptotic behavior for large values
of their argument. These  informations are required in the sequel
since the expansion coefficients of  our results for the
renormalization factors and critical exponents can be expressed in terms of
single integrals involving these functions.

In Sec.~\ref{sec:ren} we specify our renormalization procedure and present our
two-loop results for the renormalization factors.
Our $\epsilon$-expansion results for the critical, correction-to-scaling, and
crossover exponents are described in Sec.~ \ref{sec:critexp}.
Utilizing these we determine numerical estimates for the values of these exponents
in $d=3$ dimensions, which we compare with available results from Monte
Carlo calculations and other sources.
Section \ref{sec:Concl} contains a brief summary and concluding remarks.
In the Appendixes \ref{scf}--\ref{numint} various calculational details
are described.

\section {Scaling functions of the free theory and their
asymptotic behavior}\label{sec:asbeh}

\subsection{The free propagator and its scaling function}\label{ssec:freeprop}

Following the strategy utilized in I, we employ
in our perturbative renormalization scheme
the free propagator with $\tau_0=\rho_0=0$.
In position space, it is given by
\begin{equation}\label{Lprop}
G({\boldsymbol{x}})=G(x_\|,x_\perp) = \int_{\boldsymbol{q}}
{\e^{i{\boldsymbol{q}}\cdot{\boldsymbol{x}}}
\over q_\perp^2+\sigma_0{q_\|^4}}\,.
\end{equation}
Here $x_\|=|{\boldsymbol{x}}_\||$ and $x_\perp=|{\boldsymbol{x}}_\perp|$
are the Euclidean lengths of the
parallel and perpendicular components of  ${\boldsymbol{x}}$, and we
have introduced the notation
\begin{equation}\label{Inorm}
\int_{\boldsymbol{q}}\equiv\int_{{\boldsymbol{q}}_\|}
\int_{{\boldsymbol{q}}_\perp}\;\quad\mbox{with}\quad
\int_{{\boldsymbol{q}}_\|}\equiv{\int_{{\mathbb{R}}^{m}}}{\d^{m}q_\|\over (2\pi)^{m}}
\quad
\quad\mbox{and}\quad
\int_{{\boldsymbol{q}}_\perp}\equiv{\int_{{\mathbb{R}}^{d-m}}}{\d^{d-m}q_\perp\over  
(2\pi)^{d-m}}\;
\end{equation}
for integrals over momenta
$\boldsymbol{q} = ({\boldsymbol{q}}_\|,{\boldsymbol{q}}_\perp)\in
\Rset^m\times\Rset^{d-m} $. Whenever necessary, these integrals are dimensionally
regularized.

Rescaling the momenta as
${{\boldsymbol{q}}_\| }\sigma_0^{1/ 4}\sqrt{x_\perp} \to {\boldsymbol{q}}_\|$
and ${{\boldsymbol{q}}_\perp}x_\perp \to {\boldsymbol{q}}_\perp$ yields
the scaling form
\cite{DS00b} (cf.\ Ref.\ \cite{Hen97,FH93})
\begin{equation}\label{Lsprop}
G(x_\|,x_\perp)=x_\perp^{-2+\epsilon}\,\sigma_0^{-{m/ 4}}\,
{\Phi}{\left({\sigma_0^{-{1/ 4}}\,{x_\|}\, x_\perp^{-1/2}}\right)}
\end{equation}
with the scaling function
\begin{equation}\label{F}
\Phi(\upsilon)\equiv\Phi(\upsilon;m,d)= \int_{{\boldsymbol{q}}_\|}  
\int_{{\boldsymbol{q}}_\perp}
{\e^{i{{\boldsymbol{q}}_\perp}\cdot{\boldsymbol{e}}_\perp}\,
\e^{i{\boldsymbol{q}}_\|\cdot \boldsymbol{\upsilon}}\over q_\perp^2+q_\|^4}\,,
\end{equation}
where ${\boldsymbol{e}}_\perp$
is an arbitrary unit $d_\perp$-vector while $\boldsymbol{\upsilon}$ stands for the
dimensionless $d_\|$-vector
\begin{equation}
\boldsymbol{\upsilon}= \sigma_0^{-{1/ 4}}\,{\boldsymbol{x}}_\|\,x_\perp^{-1/2}\,.
\end{equation}

In I the following representation of $\Phi(\upsilon)$ in terms
of generalized hypergeometric functions was obtained:
\begin{eqnarray}
{\Phi}(\upsilon)&=&\, 2^{-2-m}\,\pi^{-{d-1\over 2}}\,\hat{\Phi}(\upsilon)\;,\\
\hat{\Phi}(\upsilon)&=&
{\Gamma{\left(1{-}{\epsilon\over 2} \right)}\over\Gamma{\left({1\over  
2}{+}{m\over 4}\right)}}
\;{\hgf}{\left(1{-}{\epsilon\over 2};
{1\over 2},{1\over 2}{+}{m\over 4};{\upsilon^4\over 64}\right)}
\nonumber\\&&\mbox{}
- {\upsilon^2\over 4}{\Gamma{\left({3\over 2}{-}{\epsilon\over 2} \right)}
\over\Gamma{\left(1{+}{m\over 4} \right)}}
\;{\hgf}{\left({3\over 2}{-}{\epsilon\over 2};
{3\over 2},1{+}{m\over 4};{\upsilon^4\over 64}\right)}\,,
\end{eqnarray}
with $\epsilon=4+\frac{m}{2}-d$.
Upon expanding the hypergeometric functions in powers of $\upsilon^4$
and resumming, one arrives at the Taylor expansion
\begin{equation}\label{sinsum}
\hat{\Phi}(\upsilon)=
\sum_{k= 0}^{\infty}{1\over k!}\,{\Gamma{\left(1-{\epsilon\over 2}+{k\over  
2}\right)}\over
\Gamma{\left({1\over 2}+{m\over 4}+{k\over 2}\right)}}\,
{\left(-{\upsilon^2\over 4} \right)}^k\;.
\end{equation}
The result tells us that $\hat\Phi$ can be written in the form
\begin{equation}\label{1psi1}
\hat\Phi(\upsilon)=
{_1\!\Psi_1}{\left[{\left(1{-}{\epsilon\over 2},{1\over 2}\right)};
{\left(\frac{1}{2}{+}\frac{m}{4},\frac{1}{2}\right)};
-{\upsilon^2\over 4}\right]}\;,
\end{equation}
where ${_1\!\Psi_1}$ is a particular one of the  Fox-Wright $\Psi$ functions
(or Wright functions) ${_p\!\Psi_q}$ \cite{Fox28,Wright35,Wright40,AM95,I-H87},
further generalizations of the generalized hypergeometric functions
$\hgpfq$ whose series representations are given by
\begin{equation}\label{pPq}
{\ppsiq}[\{(a_i,A_i)\},\{(b_j,B_j\};x]=
\sum_{k=0}^\infty{1\over k!}\;
{\prod_{i=1}^p\Gamma(a_i+A_i\, k)\over\prod_{i=1}^q
\Gamma(b_j+B_j\, k)}\;x^k\;.
\end{equation}

In the sequel, we shall need the asymptotic behavior of $\Phi(\upsilon)$ as
$\upsilon\to\infty$. This may be inferred from theorems due to Wright
\cite{Wright35,Wright40} about the asymptotic expansions
of the functions $\ppsiq$. We discuss this matter in Appendix \ref{scf},
where we show that the asymptotic expansion these theorems predict
for nonexceptional values of $m$ and $\epsilon$,
\begin{equation}\label{asexpPhihat}
\hat{\Phi}(\upsilon\to\infty)\approx{\left({\upsilon\over  
2}\right)}^{-4+2\epsilon}\sum_{k=0}^\infty
{(-1)^k\over k!}\,
{2\,\Gamma(2-\epsilon+2k)\over\Gamma({m\over 4}{-}{1\over 2}{+}{\epsilon\over  
2}{-}k)}\,
{\left({\upsilon\over 2} \right)}^{-4k}\;,
\end{equation}
follows from the integral representation (\ref{F}) in an equally straightforward
manner as the Taylor expansion (\ref{sinsum}). Nonexceptional values
of $m$ and $\epsilon$ are characterized by the property that
none of the poles which the nominator of the coefficient
\begin{equation}
f(k)={\Gamma{\left(1-{\epsilon\over 2}+{k\over 2}\right)}\over
\Gamma{\left({1\over 2}+{m\over 4}+{k\over 2}\right)}}
\end{equation}
of the power series (\ref{sinsum}) has
at
\begin{equation}
k=k_l\equiv \epsilon-2\,l\;,\quad l=1,2,\ldots,
\end{equation}
gets canceled
by a pole of the denominator. If $\epsilon=0$, the only values
among $m=1,2,\ldots,7$ for which such cancellations  occur are
$m=2$ and $m=6$. More generally, this happens for $d=m+1$ and
$d=m+3$ where the expansion (\ref{asexpPhihat}) terminates
after the first ($k=0$) term and vanishes identically, respectively.
In accordance with Wright's theorems, corrections to these
truncated expansions are exponentially small.
In fact, in these two cases $\Phi(\upsilon)$ reduces
to the much simpler expressions
\begin{equation}\label{Phidm+1}
\Phi(\upsilon)={u^{2 - m}\over 8\,{\pi}^{m/2}}
\;{\gamma}{\left(\frac{m-2}{2},\frac{\upsilon^2}{4}\right)}\;,
\qquad d=m+1\,,
\end{equation}
and
\begin{equation}\label{Phidm+3}
\Phi(\upsilon)={(4\,\pi)^{-{m+2\over 2}}}\,
\e^{-{\upsilon^2/ 4}}\;,\qquad d=m+3\,,
\end{equation}
where $\gamma(a,x)$ is the incomplete gamma function.
These equations comprise two cases where $d$ becomes the upper critical
dimension (\ref{ucd}), namely $(d,m)=(6,7)$ and $(d,m)=(2,5)$.
In the former, Eq.~(\ref{Phidm+1}) simplifies to
\begin{equation}\label{Phid*6}
\Phi(\upsilon)={1\over (2\pi)^3}\,{1\over  
\upsilon^4}\,{\left[1-{\left(1+{\upsilon^2\over 4}\right)}\,
{\e}^{-{\upsilon^2/ 4}}\right]}\;,\quad m=6\,,\;d=d^*=7\,.
\end{equation}
This  result as well as
Eq.~(\ref{Phidm+3}) with $m=2$ were employed in I,
 where we also derived
the leading term ($k=0$) of the asymptotic series (\ref{asexpPhihat}).

For general values of $m$ and $d=d^*$, the scaling function
$\Phi(\upsilon;m,d)\equiv\Phi(\upsilon)$ can be written as
\begin{eqnarray}\label{Phid*}
{\Phi}(\upsilon;m,d^*)=
\frac{1}{2^{2+m}\,\pi^{\frac{6{+}m}{4}}}\,{\left[
{{\hgf}{\left( 1;{1\over 2},{2{+}m\over 4};
{\upsilon^4\over 64}\right)}\over\Gamma{\left({2{+}m\over 4} \right)}}
 -\sqrt{\pi}\,{\left({{\upsilon}^2\over 8}\right)}^{1{-}{m\over 4}} \;
{{I}_{m\over 4}}{\left({\upsilon^2\over 4}\right)}\right]}
\nonumber\\
=\frac{1}{2^{2+m}\,\pi^{\frac{6{+}m}{4}}}\,{\left\{
\frac{1}{\Gamma{\left(\frac{2+m}{4}\right)}}+
\sqrt{\pi}\,{\left({{\upsilon}^2\over 8}\right)}^{1{-}{m\over 4}}{\left[
{\modStruveL}_{\frac{m}{4}}{\left({\upsilon^2\over 4}\right)}-
{{I}_{m\over 4}}{\left({\upsilon^2\over 4}\right)}\right]}
\right\}},
\nonumber\\
\end{eqnarray}
where $\modStruveL_\alpha(z)$ and $I_\alpha(z)$ are
modified Struve and Bessel functions,
respectively \cite{AS72}.

The second form is in conformity with the one
given by Frachebourg and Henkel \cite{FH93} for the case $m=1$.
These authors encountered this (and similar) scaling functions
when studying Lifshitz points of order $L-1$ of spherical models.
They also analyzed the large-$\upsilon$ behavior of these functions,
verifying explicitly the asymptotic forms predicted by Wright's
theorems. If we let $a=1$ and set $x=\upsilon^2/4$,
their scaling function denoted $\Psi(a,x)$
corresponds precisely to our $\hat\Phi(\upsilon)$,
and the asymptotic expansion
they found is consistent
with ours in  Eq.~(\ref{asexpPhihat}) and the large-$\upsilon$ form
(\ref{asPhid*}) presented below.%
\footnote{Note that the Hamiltonians of the spherical models
considered in Ref.~\cite{FH93}
involve instead of ${(\Lapar
 \boldsymbol{\phi})}^2$ a derivative term of the form
$\sum_{i=1}^m(\partial^2_i{\boldsymbol{\phi}})^2$, which breaks
the rotational invariance in the parallel subspace $\Rset^m$. Comparisons
with our results for the free theory are therefore only possible
for $L=2$ and $m=1$.}

Finally, let us explicitly give the large-$\upsilon$
forms of $\Phi(\upsilon,m,d^*)$ as implied by Eq.~(\ref{asexpPhihat}):
\begin{eqnarray}\label{asPhid*}
\Phi(\upsilon;m,d^*)&\mathop{\approx}\limits_{\upsilon\to\infty}&
 2^{1-m}\,\pi^{-{6+m\over 4}}\,
{m-2\over \Gamma({m+2\over  4})}
{\left[{1\over \upsilon^4}-{24\,(m-6)\over \upsilon^8}\right.}
\nonumber\\[0.5em]
&&{\qquad\qquad\qquad\left.\mbox{}
+{960\,(m-10)(m-6)\over\upsilon^{12}}
+O{\left(\upsilon^{-16}\right)}
\right]}.
\end{eqnarray}
In accordance with our previous considerations,
all terms or all but the first one of this series
vanish when $m=2$ or $m=6$, respectively.

\subsection{Other required scaling functions of the free  
theory}\label{ssub:othfreescf}

Besides  $\Phi$, our results to be given below
involve two other scaling functions. One is the function $\Xi(\upsilon;m,d)$
of I. This is defined through
\begin{equation}\label{Xiintro}
-(\nabla_\| G*{\nabla_\|}G)(\boldsymbol{x})=x_\perp^{-1+\epsilon}\,
\sigma_0^{-(m+2)/4}\,
\Xi\big(\sigma_0^{-1/4}x_\parallel x_\perp^{-1/2}\big)\,,
\end{equation}
where the asterisk indicates a convolution, i.e.,  
$(f{*}g)({\boldsymbol{x}})\equiv  
{\int}\d^dx'\,f({\boldsymbol{x}}-{\boldsymbol{x}}')\,g({\boldsymbol{x}}')$. The
explicit form of $\Xi(\upsilon;m,d^*)$ may be gleaned from
Eq.~(A5) of I, where it was given
in terms of Bessel and hypergeometric functions.
This can be written more compactly as
\begin{equation}\label{Xid*}
\Xi(\upsilon;m,d^*)=\frac{{\upsilon}^{2{-}{m\over  
2}}}{2^{6+\frac{m}{4}}\,\pi^{\frac{4{+}m}{4}}}\,
{\left[
{I}_{\frac{m-4}{4}}{\left({\upsilon^2\over 4}\right)}-
{{\modStruveL}_{m-4\over 4}}{\left({\upsilon^2\over 4}\right)}\right]}\;.
\end{equation}
The asymptotic expansion of the difference of functions in the square brackets
of this equations follows from
 Eqs.~(12.2.6) and (9.7.1) of Ref.~\cite{AS72}, implying
\begin{eqnarray}\label{asexpXid*}
\Xi(\upsilon;m,d^*)&\mathop{\approx}\limits_{\upsilon\to\infty}&
2^{-2-m}\,\pi^{-\frac{6 + m}{4}}  
{m-2\over\Gamma{\left(\frac{m+2}{4}\right)}}\,\upsilon^{-2}\,{\left[1+{6-m\over  
2}\,{4^2\over \upsilon^4}\right.}
\nonumber\\&&\mbox{}+{\left.
{6-m\over 2}\,{3\,(10-m)\over 2}\,{4^4\over \upsilon^8}
+O{\left(\upsilon^{-12}\right)}
\right]}\;.
\end{eqnarray}
The leading term $\sim \upsilon^{-2}$ was already given in I.
Note also that again all terms or all but the first one of this series
vanish when $m=2$ or $m=6$, respectively,
as is borne out by the explicit forms
\begin{equation}
\Xi(\upsilon;2,5)=\frac{1}{2}\,\Phi(\upsilon;2,5)=
\frac{\e^{-\upsilon^2/4}}{32\,\pi^2}
\end{equation}
and
\begin{equation}
\Xi(\upsilon;6,7)={1-\e^{-\upsilon^2/4}\over (4\,\pi)^{3}\, \upsilon^{2}}\;.
\end{equation}

The third scaling function we shall need is defined through the
Taylor expansion
\begin{equation}\label{Theta}
\Theta(\upsilon;m)\equiv \sum_{k=1}^\infty{\Gamma{\left({k\over 2}\right)}
\over k!\;\Gamma{\left({1\over 2}{+}{m\over 4}{+}{k\over 2}\right)}}\,
{\left({-\upsilon^2\over 4}\right)}^{k}\;.
\end{equation}
It can be expressed in terms of generalized hypergeometric functions $\hgpfq$
by summing the contributions with even and odd $k$ separately.
One finds
\begin{eqnarray}\label{simf}
\Theta(\upsilon;m)&=&{\upsilon^4\over 32}\,
{1\over\Gamma({3\over 2}{+}{m\over 4})}
\,\hgtwofthree{\left(1,1;{3\over 2},2,{3\over 2}{+}{m\over 4};
{\upsilon^4\over 64}\right)}
\nonumber\\&&\mbox{}
-{\upsilon^2\over 4}{\sqrt{\pi}\over\Gamma(1{+}{m\over 4})}
\;\hgf{\left({1\over 2};{3\over 2},1{+}{m\over 4};{\upsilon^4\over 64}\right)}\;.
\end{eqnarray}
Details of how this function arises in the
computation of the Laurent expansion of the four-point graph
\raisebox{-4pt}{\begin{texdraw}
\drawdim pt \setunitscale 1.5   \linewd 0.3
\move(-7 -1.5)\rlvec(15 6)
\move(-7 1.5)\rlvec(15 -6)
\move(5 0)
\lellip rx:1 ry:3
\end{texdraw}}
may be found in Appendix \ref{fourptgr}. In Appendix \ref{asbehTheta}
we show that
$\Theta(\upsilon,m)$ behaves as
\begin{eqnarray}\label{eq:asbehTheta}
\Theta(\upsilon;m)&\mathop{\approx}\limits_{\upsilon\to\infty}&{1\over  
\Gamma{\left({2+m\over 4}\right)}}\,{\left[
-\ln{\upsilon^4\over 16}+
\psi{\left({m+2\over 4}\right)}
-C_E
-8\,\frac{m-2}{\upsilon^4}\right.}
\nonumber\\&&\qquad\qquad\qquad{\left.\mbox{}
+96\, \frac{(m-2)(m-6)}{\upsilon^8}
\right]}
+O{\left(\upsilon^{-12}\right)}
\end{eqnarray}
in the large-$\upsilon$ limit, where $C_E\simeq 0.577216$ is
Euler's constant and while
$\psi(x)$ denotes the digamma function.

For the special values $m=2$ and $m=6$, the function $\Theta(\upsilon;m)$ reduces to
a sum of elementary functions and the exponential integral function $E_1(x)$.
As can be easily deduced from the series expansion (\ref{Theta}), one has
\begin{equation}
\Theta(\upsilon,2)=-2\,{\left[C_E+\ln{\upsilon^2\over 4}+
E_1{\left({\upsilon^2\over 4}\right)}\right]}
\end{equation}
and
\begin{eqnarray}
\Theta(\upsilon,6)&=&1-2\,C_E-\ln{\upsilon^4\over  
16}-2\,E_1{\left({\upsilon^2\over 4}\right)}+\frac{8\,\e^{-{\upsilon^2/  
4}}}{\upsilon^2}+32\,\frac{\e^{-{\upsilon^2/ 4}}-1}{\upsilon^4}\nonumber\\
&=&1+\Theta(\upsilon;2)-\hat{\Phi}(\upsilon;6,7)\;,
\end{eqnarray}
respectively.

\section{Renormalization}\label{sec:ren}
\subsection{Reparametrizations}\label{ssec:repara}
>From I we know that the ultraviolet singularities
of the $N$-point correlation functions  
$\langle{\prod_{\nu=1}^N}\phi_{i_\nu}({\boldsymbol{x}}_\nu )\rangle$
of the Hamiltonian (\ref{Ham}) can be absorbed via reparametrizations of the form
\begin{equation}
\phi={Z_\phi}^{{1}/{2}}\,\phi_{\mathrm{ren}}\;,
\end{equation}
\begin{equation}
\tau_0-\tau_{0c}=\mu^2\,Z_\tau\,\tau\;,
\end{equation}
\begin{equation}
\sigma_0=Z_\sigma\,\sigma\;,
\end{equation}
\begin{equation}
u_0\,{\sigma_0}^{-m/4}\,F_{m,\epsilon}=\mu^\epsilon\,Z_u\,u\;,
\end{equation}
and
\begin{equation}
\left(\rho_0-\rho_{0c}\right)\,{\sigma_0}^{-1/2}
=\mu\,Z_\rho\,\rho\;.
\end{equation}
Here $\mu$ is an arbitrary momentum scale. The critical values
of the Lifshitz point, $\sigma_{0c}$ and
$\rho_{0c}$, vanish in our perturbative RG scheme based on dimensional
regularization and the $\epsilon$ expansion. The factor $F_{m,\epsilon}$
serves to choose a convenient normalization of $u$. A useful choice is to
write the following one-loop integral for
\raisebox{-2.2pt}{\begin{texdraw}
\drawdim pt \setunitscale 2.5   \linewd 0.3
\lellip rx:4 ry:1.8
\move(4 0)\rlvec(2 2)
\move(4 0)\rlvec(2 -2)
\move(-4 0)\rlvec(-2 -2)
\move(-4 0)\rlvec(-2 2)
\end{texdraw}}
as
\begin{equation}\label{I2e}
{\int_{\boldsymbol{q}}}{1\over (q_\|^4+q_\perp^2)[
q_\|^4+({\boldsymbol{q}}_\perp+{\boldsymbol{e}}_\perp)^2]}=
\frac{F_{m,\epsilon}}{\epsilon}\;.
\end{equation}
This integral is evaluated in Appendix \ref{fourptgr}.
The result, given in Eq.~(\ref{I2}), yields
\begin{eqnarray}\label{Fmeps}
F_{m,\epsilon}&=&(4\,\pi)^{-{8+m-2\,\epsilon\over 4}}\,
\frac{\Gamma{\left(1+{\epsilon\over 2}\right)}
\,\Gamma^2{\left(1-{\epsilon\over  
2}\right)}\,\Gamma{\left(\frac{m}{4}\right)}}{\Gamma(2-\epsilon)\,
\Gamma{\left(\frac{m}{2}\right)}}\nonumber\\
&=&\frac{(4\, \pi)^{-2 - \frac{m}{4}}\,
    \Gamma(\frac{m}{4})}{\Gamma(\frac{m}{2})}\,{\left[1+
(2-C_E+\ln4\pi)\,{\epsilon\over 2}+O(\epsilon^2)
\right]}\;.
\end{eqnarray}

\subsection{Renormalization factors}\label{ssec:renfac}

With this different choice of normalization of the coupling constant,
our results of I for $Z_\phi$, $Z_\sigma$, and $Z_\rho$
translate into
\begin{equation}\label{Zphi}
Z_\phi=1-\ntwo\, \frac{j_\phi(m)}{12\,(8-m)}\,\frac{u^2}{\epsilon}+O(u^3)\;,
\end{equation}
\begin{equation}\label{Zsigma}
{Z_\sigma}Z_\phi=1+\ntwo\,{j_{\sigma}(m)\over 96\,m(m{+}2)}
\,\frac{u^2}{\epsilon}+O(u^3)\;,
\end{equation}
and
\begin{equation}\label{Zrho}
{Z_\rho}{Z_\phi}{Z_\sigma^{1/2}}=1+\ntwo\,{j_{\rho}(m)\over 8\,m}
\,\frac{u^2}{\epsilon}+O(u^3)
\end{equation}
with
\begin{equation}\label{jphidef}
j_\phi(m)\equiv B_m
\,{\int_0^\infty}\!{\d}\upsilon\,\upsilon^{m-1}\,\Phi^3(\upsilon;m,d^*)\;,
\end{equation}
\begin{equation}\label{jsigmadef}
j_\sigma(m)\equiv  
B_m\,{\int_0^\infty}\!{\d}\upsilon\,\upsilon^{m+3}\,\Phi^3(\upsilon;m,d^*)\;,
\end{equation}
and
\begin{equation}\label{jrhodef}
j_\rho(m)\equiv  
B_m\,{\int_0^\infty}\!{\d}\upsilon\,\upsilon^{m+1}\,\Phi^2(\upsilon;m,d^*)\,
\Xi(\upsilon;m,d^*)\;.
\end{equation}
Except for the factor $B_m$, which is
\begin{equation}\label{Bm}
B_m\equiv \frac{S_{4-{m\over 2}}\, S_m}{F_{m,0}^2}=
{\frac{{2^{10 + m}}\,{{\pi }^{6 + {\frac{3\,m}{4}}}}\,
     \Gamma({\frac{m}{2}})}{\Gamma(
      2 - {\frac{m}{4}})\,
     {{\Gamma({\frac{m}{4}})}^2}}}\;,
\end{equation}
the  coefficients $j_\phi(m)$, $j_\sigma(m)$, and $j_\rho(m)$
are precisely the integrals denoted respectively as
$J_{0,3}(m,d^*)$, $J_{4,3}(m,d^*)$, and $I_1(m,d^*)$,
in I. The quantity $S_d=2\,\pi^{d/2}/\Gamma(d/2)$
in Eq.~(\ref{Bm}) (with $d=m$, e.g.)
means the surface area of a $d$-dimensional unit sphere.

Our two-loop results for the remaining $Z$-factors, $Z_u$ and $Z_\tau$,
can be written as
\begin{equation}\label{Ztau}
{Z_\tau}Z_\phi=1+\frac{n{+}2}{3} \,\frac{u}{2\,\epsilon}
+\frac{n{+}2}{3} {\left[\nfive\,\frac{1}{\epsilon^2}-{J_u(m)\over 2\,\epsilon}
\right]}
\,{u^2\over 2}+O(u^3)
\end{equation}
and
\begin{eqnarray}\label{Zu}
{Z_u}{Z_\phi^2}{Z_\sigma^{m/4}}&=&1+\neight
\,\frac{3\,u}{2\,\epsilon}+{\left[{\left(\neight\,{3\over 2\,\epsilon}\right)}^2
-3\,{5n{+}22\over 27}\,
{J_u(m)\over 2\,\epsilon}
\right]}\,u^2
\nonumber\\&&
\mbox{}+O(u^3)\;,
\end{eqnarray}
with
\begin{equation}\label{Ju}
J_u(m)=1-{C_E+\psi{\left(2-\frac{m}{4}\right)}\over 2}+
j_u(m)\;,
\end{equation}
where $j_u(m)$ means the integral
\begin{equation}\label{judef}
j_u(m)=\frac{B_m}{2^{4+m}\,\pi^{(6+m)/4}}\,
{\int_0^\infty}\d\upsilon\,\upsilon^{m-1}\,\Phi^2(\upsilon;m,d^*)\,
\Theta(\upsilon;m)\;.
\end{equation}

For the values $m=2$ and $m=6$, the above integrals $j_\phi$, $j_\sigma$,
$j_\rho$, and $j_u$ can be computed
analytically (cf.\ I and Appendices \ref{Lexp} and \ref{fourptgr}).
This gives
\begin{equation}\label{jphi26}
j_\phi(2)=\frac{4}{3}\;,\qquad  
j_\phi(6)=\frac{8}{3}\,{\left[1-3\,\ln{4\over3}\right]}\;,
\end{equation}
\begin{equation}\label{jsigma26}
j_\sigma(2)=\frac{128}{27}\;,\qquad j_\sigma(6)=\frac{448}{9}\;,
\end{equation}
\begin{equation}\label{jrho26}
j_\rho(2)=\frac{8}{9}\;,\qquad  
j_\rho(6)=\frac{8}{3}\,{\left[1+6\ln{4\over3}\right]}\;,
\end{equation}
\begin{equation}\label{ju26}
j_u(2)=-\ln\frac{3}{2}\;,\qquad j_u(6)=
-{\left(\frac{1}{6}+\ln{128\over 27}\right)}\;,
\end{equation}
and
\begin{equation}\label{Ju26}
J_u(2)=\ln\frac{4}{3}\;,\qquad J_u(6)=\frac{5}{6}-3\,\ln\frac{4}{3}\;.
\end{equation}

For other values of $m$ we determined these integrals by numerical integration
in the manner explained in Appendix \ref{numint}. The results are listed in  
Table \ref{tab:jvalues}.\medskip

\begin{table}[htb]
\caption{Numerical values of the integrals $j_\phi(m)$, \ldots,
$J_u(m)$.}\label{tab:jvalues}
\begin{tabular}{cccccccc}\hline
m&1&2&3&4&5&6&7\\\hline
$j_\phi$&1.642(9)&1.33333&1.055(6)&0.803(7)&0.57(4)&0.36521&0.17(4)\\
$j_\sigma$&1.339(4)&4.74074&\hspace*{-1ex}10.804(3)&
\hspace*{-1ex}20.067(7)&\hspace*{-1ex}32.95(4)&
\hspace*{-1ex}49.77778&\hspace*{-1ex}70.74(7)\\
$j_\rho$&0.190(6)&0.88889&1.999(9)&3.464(1)&5.23(4)&7.26958&9.53(6)\\
$j_u$&\hspace*{-1ex}-0.203(7)&\hspace*{-1ex}-0.40547&\hspace*{-1ex}-0.624(2)
&\hspace*{-1ex}-0.880(1)&\hspace*{-1ex}-1.21(1)&
\hspace*{-1ex}-1.72286&\hspace*{-1ex}-2.92(4)\\
$J_u$&0.383(8)&0.28768&0.200(8)&0.119(8)&0.04(3)&\hspace*{-1ex}-0.02971&
\hspace*{-1ex}-0.09(9)\\\hline
\end{tabular}
\end{table}

\subsection{Beta function and fixed-point value $u^*$}\label{ssec:betaf}

>From the above results for the renormalization factors the
beta function
\begin{equation}
\beta_u(u)\equiv{\left.\mu\partial_\mu\right|_0}\,u
\end{equation}
and the exponent functions
\begin{equation}
\eta_\iota\equiv{\left.\mu\partial_\mu\right|_0}\,\ln Z_\iota\,,\quad
\iota=\phi,\,\sigma,\,\rho,\,\tau,\,u,
\end{equation}
can be calculated in a straightforward manner.
 Here ${\left.\partial_\mu\right|_0}$
denotes a derivative at fixed bare values $\sigma_0$, $\tau_0$, $\rho_0$, and $u_0$.
Since we employed minimal subtraction of poles, the exponent functions
satisfy the following simple relationship to the residua ($\res$) of the $Z$ factors:
\begin{equation}\label{etaiota}
\eta_\iota(u)=-u\partial_u\res_{\epsilon=0}[Z_\iota(u)]\;,\quad
\iota=\phi,\,\sigma,\,\rho,\,\tau,\,u.
\end{equation}

For the beta function, which is related to $\eta_u$ via  
$\beta_u(u)=-u[\epsilon+\eta_u(u)]$, we obtain
\begin{eqnarray}\label{betau}
\beta_u(u)&=&-\epsilon\,u+\neight\,\frac{3}{2}\,u^2
-{\Bigg\{3\,{5n{+}22\over 27}\,J_u(m)}
\nonumber\\&&\mbox{}+{\left.\frac{1}{24}\,\ntwo{\left[
{j_{\sigma}(m)\over 8\,(m{+}2)}-
j_\phi(m)
\right]}
\right\}}\,u^3+O(u^4)\;.
\end{eqnarray}
Upon solving for the nontrivial zero of $\beta_u$, we see that the infrared-stable
fixed point is located at
\begin{eqnarray}\label{ustar}
u^*&=&\frac{2\,\epsilon}{3}\,\frac{9}{n{+}8}+\frac{8\,\epsilon^2}{27}\,
{\left(\frac{9}{n{+}8}\right)}^3\,{\Bigg\{
3\,{5n{+}22\over 27}\,J_u(m)}\nonumber\\
&&\qquad\qquad\quad\mbox{}+{\left.\frac{1}{24}\,\ntwo
\,{\left[{j_{\sigma}(m)\over 8\,(m{+}2)}-j_\phi(m)
\right]}
\right\}}+O{(\epsilon^3)}\;.
\end{eqnarray}

We refrain from giving the resulting lengthy expressions for the exponent functions
here. The values $\eta_\iota^*\equiv\eta_\iota(u^*)$ of these functions at the  
infrared-stable fixed point are presented in  
Eqs.~(\ref{etal2})--(\ref{thetaeps}) below.

\section{Critical exponents}\label{sec:critexp}

\subsection{Analytic $\epsilon$-expansion results}\label{ssec:epsexp}

The fixed-point value (\ref{ustar}) can now be substituted
into the exponent functions (\ref{etaiota}) that are implied by our results
(\ref{Zphi})--(\ref{Zrho}) and (\ref{Ztau}) for the renormalization factors
to obtain the universal quantities $\eta_\iota^*=
\eta_\iota(u^*)$, $\iota=\phi$, $\sigma$, $\rho$, and $\tau$.
Recalling how these are related to the critical exponents
(cf.\ I), one arrives at the $\epsilon$ expansions
\begin{equation}\label{etal2}
\eta_{l2}=\eta_\phi^*=\frac{n+2}{(n+8)^2}\,
\frac{2\,j_\phi(m) }{8- m} \,\epsilon^2+O(\epsilon^3)\;,
\end{equation}
\begin{equation}\label{etal4}
\eta_{l4}=4 \,\frac{\eta_\phi^*+\eta_\sigma^*}{2+\eta_\sigma^*}
=-\frac{n+2}{(n+8)^2}\,
\frac{j_\sigma(m) }{2m\,(m+2)} \,\epsilon^2+O(\epsilon^3)\;,
\end{equation}
\begin{eqnarray}\label{etataueps}
{1\over\nu_{l2}}-2=\eta_\tau^*&=&-{n{+}2\over n{+}8}\,{\epsilon}
-{n{+}2\over 2\,(n{+}8)^2}\Bigg\{4\,{7n{+}20\over n{+}8}J_u(m)
+{n{+}2\over n{+}8}\,{j_\sigma(m)\over 8\,(m{+}2)}
\nonumber\\
&&\mbox{}
+{m\,(n{+}2)+4\,(4{-}n)\over(n{+}8)(8{-}m)}\,j_\phi(m)
\Bigg\}\epsilon^2+O(\epsilon^3)\,,
\end{eqnarray}
\begin{eqnarray}\label{nul4eps}
\nu_{l4}=\frac{2+\eta_\sigma^*}{4\,(2+\eta_\tau^*)}&=&\frac{1}{4}+{n{+}2\over  
n{+}8}\,{\epsilon\over 8}+
{1\over 16}\,{n+2\over (n{+}8)^2}
\Bigg\{ n{+}2+4\,\frac{7n{+}20}{ n{+}8}\, J_u(m)
\nonumber\\
&&{\mbox{}-{n{+}2\over n{+}8}\,j_\phi(m)
-{\left[1-{m\over 4}\,{n{+}2\over n{+}8}\right]}\,{j_\sigma(m)\over 2m(m{+}2)}
\Bigg\}}\epsilon^2+O(\epsilon^3)\;,\nonumber\\
\end{eqnarray}
and
\begin{equation}\label{etarhoeps}
{\varphi\over\nu_{l2}}-1=\eta_\rho^*={n+2\over  
(n+8)^2}\,{\left[{j_\sigma(m)\over  
8\,m\,(m{+}2)}-\frac{3\,j_\rho(m)}{m}-\frac{j_\phi(m)}{8{-}m}
\right]}\,\epsilon^2+O(\epsilon^3)\,.
\end{equation}

The anisotropy exponent $\theta={\nu_{l4}/\nu_{l2}}$
of Eq.~(\ref{asci}) is given by
\begin{equation}\label{thetaeps}
\theta=\frac{2+\eta_\sigma^*}{4}=\frac{1}{2}-
\frac{n+2}{2\,(n+8)^2}\,{\left[
{j_\sigma(m)\over 8\,m\,(m+2)}+{j_\phi(m)\over 8-m}
\right]}\,\epsilon^2+O(\epsilon^3)\;,
\end{equation}
and for the correction-to-scaling exponent, we obtain
\begin{eqnarray}\label{omega}
\omega_{l2}\equiv\beta_u'(u^*)&=&
\epsilon-\frac{36\,\epsilon^2}{(n+8)^2}\,
{\left\{3\,{5n{+}22\over 27}\,J_u(m)\right.}
\nonumber\\&&\qquad\qquad\mbox{}+{\left.\frac{1}{24}\,\ntwo{\left[
{j_{\sigma}(m)\over 8\,(m{+}2)}-
j_\phi(m)
\right]}
\right\}}+O(\epsilon^3)\;.
\end{eqnarray}
Our rationale for denoting the latter analog of  the usual Wegner exponent
as $\omega_{l2}$ is the following: It governs those corrections to scaling
that are weaker by a factor of
$\xi_\perp^{-\omega_{l2}}\sim |\tau|^{\nu_{l2}\,\omega_{l2}}$ than the leading
infrared singularities. Since  
$\xi_\perp^{-\omega_{l2}}\sim\xi_\|^{-{\omega_{l2}/\theta}}$
it is natural to introduce also the related correction-to-scaling exponent
\begin{equation}\label{omegal4def}
\omega_{l4}\equiv {\omega_{l2}\over\theta}\,.
\end{equation}
In the case of an isotropic Lifshitz point (cf.\ Sec.~\ref{ssec:isoLP}),
in which only the correlation length $\xi_\|$ is left, this exponent
retains its significance and becomes the sole remaining analog of  Wegner's exponent.

\subsection{The special cases $m=2$ and $m=6$}\label{sec:scm26}

For these two special cases, a two-loop calculation was performed in
Ref.~\cite{MC99}. In order to compare its results with ours, we must
recall that these authors took $d_\|=m-\epsilon_\|$
and  $d_\perp=4-m/2$, with $m=2$ and $m=6$. Accordingly, our $\epsilon$
must be identified with $\epsilon= 2\,\epsilon_\|$.
If we substitute the analytic values (\ref{jphi26})--(\ref{Ju26})
of the integrals $j_\phi,\ldots,J_u$ into our results (\ref{etal2})--(\ref{nul4eps}),
the latter reduce to
\begin{equation}\label{etal22}
\eta_{l2}(m{=}2)=\frac{4}{9}\,\frac{n+2}{(n+8)^2}\,\epsilon^2+O(\epsilon^3)\;,
\end{equation}
\begin{equation}\label{etal42}
\eta_{l4}(m{=}2)=-\frac{8}{27}\,\frac{n+2}{(n+8)^2}\,\epsilon^2+O(\epsilon^3)\;,
\end{equation}
\begin{eqnarray}\label{nul22}
\nu_{l2}(m{=}2)&=&\frac{1}{2}+\frac{n{+}2}{4\,(n{+}8)}\,\epsilon+
\frac{2\,(n{+}2 )}{( n{+}8)^3}\,\bigg[ {n^2\over 16} + {131\,n\over 216} +  
{35\over 27}
  \nonumber\\&&      +{7\,n  {+} 20 \over 4}
          \,\ln{\frac{4}{3}}
        \bigg]\epsilon^2 +O(\epsilon^3)\;,
\end{eqnarray}
\begin{eqnarray}\label{nul42}
\nu_{l4}(m{=}2)&=&\frac{1}{4}+{n{+}2\over n{+}8}\,{\epsilon\over  
8}+\frac{n{+}2}{(n{+}8)^3}
\bigg[{n^2\over 16} + {115\,n\over 216} + {19\over 27}
\nonumber\\&&
\mbox{}
 + {7\,n  {+} 20\over 4} \,\ln {\frac{4}{3}}\bigg]\,\epsilon^2+O(\epsilon^3)\;,
\end{eqnarray}
and
\begin{equation}\label{etal26}
\eta_{l2}(m{=}6)=8\,{\left(\frac{1}{3}-\ln\frac{4}{3}\right)}
\,\frac{n+2}{(n+8)^2}\,\epsilon^2+O(\epsilon^3)\;,
\end{equation}
\begin{equation}\label{etal46}
\eta_{l4}(m{=}6)=-\frac{14}{27}
\,\frac{n+2}{(n+8)^2}\,\epsilon^2+O(\epsilon^3)\;,
\end{equation}
\begin{eqnarray}\label{nul26}
\nu_{l2}(m{=}6)&=&\frac{1}{2}+\frac{n{+}2}{4\,(n{+}8)}\,\epsilon+
\frac{2\,(n{+}2) }{( n{+}8)^3}\,\bigg[ {n^2\over 16} + {331\,n\over 144} +  
{547\over 72}
  \nonumber\\&&      - {23\, n{+}88\over 4}
          \,\ln{\frac{4}{3}}
        \bigg]\epsilon^2 +O(\epsilon^3)\;,
\end{eqnarray}
\begin{eqnarray}\label{nul46}
\nu_{l4}(m{=}6)&=&\frac{1}{4}+{n{+}2\over n{+}8}\,{\epsilon\over  
8}+\frac{n{+}2}{4(n{+}8)^3}
\bigg[{n^2\over 4} + {835\,n\over 108} + {1009\over 54}
\nonumber\\&&
\mbox{}
 -( 19\,n  {+} 56) \,\ln {\frac{4}{3}}\bigg]\epsilon^2+O(\epsilon^3)\;,
\end{eqnarray}
respectively. Our results (\ref{etal22}), (\ref{etal42}), (\ref{etal26}), and  
(\ref{etal46}) for the correlation exponents are consistent with Eqs.~(57),  
(56),
(61), and (60) of Ref.~\cite{MC99}. However,
the $\epsilon^2$ terms of the correlation-length exponents given in its
Eqs.~(58), (59), (62), and (63) are incompatible with ours.
These discrepancies have two causes. There is a sign error
in  Mergulh{\~a}o and Carneiro's \cite{MC99} definition (50):
The term in its second line should be
replaced by its negative; only then are their
general formulas (54) and (55) for the correlation-length exponents
$\nu_{l\alpha}$ ($\equiv$ our $\nu_{l4}$) and $\nu_{l\beta}$ ($\equiv $ our  
$\nu_{l2}$) correct.
With this correction, these formulas yield results in conformity with ours if  $m=6$.
However, in the case $m=2$, another correction must be made:
We believe that in their Eq.~(C7) for the integral $I_3$ the prefactor
of the multiple integral  on the right-hand side is too small by a
 factor of $2^{\epsilon}$. This entails that
the logarithm term $\ln (16/ 3)$ of the
$\epsilon^{-1}$ pole in their Eq.~(C10) gets modified to
$\ln (64/ 3)$. With this additional correction, the $\epsilon^2$ terms
following for $m=2$ from their corrected
Eqs.~(54) and (55) turn out to be consistent with
ours.\footnote{We are grateful to C. E. I. Carneiro who checked the calculations
of Ref. \cite{MC99} and confirmed the correctness of our results.}

\subsection{Numerical values of the second-order expansion coefficients}

In order to analyze further the above results, let us denote
the coefficients of the $\epsilon^2$ terms of the exponents  
$\lambda=\nu_{l2}$, $\nu_{l4}$,\ldots, $\varphi$ as
$C_2^{(\lambda)}(n,m)$, so that, for example,
\begin{equation}
\nu_{l2}={1\over 2}+\frac{1}{4}\,{n{+}2\over n{+}8}\,\epsilon
+C_2^{(\nu_{l2})}(n,m)\,\epsilon^2
+O(\epsilon^3)\,.
\end{equation}
The numerical values of these coefficients for $m=1,\ldots,6$ are listed in  
Table \ref{tab:C2n=1} for the case $n=1$.

\begin{table}[htb]
\caption{Numerical values of the second-order expansion
coefficients $C^{(\lambda)}_2(n{=}1,m)$
\label{tab:C2n=1}}
\begin{tabular}{cccccccc}\hline
$m$&0&1&2&3&4&5&6\\\hline
$C_2^{(\nu_{l2})}$& 0.043 & 0.037113 & 0.032158 & 0.027744 &
 0.023677& 0.01987 & 0.01626 \\
$C_2^{(\nu_{l4})}$& --- & 0.015867 & 0.013335 & 0.011083 & 0.009010
                           & 0.00707 & 0.00524 \\
$C_2^{(\alpha_l)}$&\hspace*{-1ex}-0.090 &\hspace*{-1ex}-0.062429
&\hspace*{-1ex}-0.039810 &\hspace*{-1ex}-0.019277
&\hspace*{-1ex}-0.000059 & 0.01817 & 0.03564\\
$C_2^{(\beta_l)}$& 0.006 & \hspace*{-1ex}-0.00155\phantom{0}&
\hspace*{-1ex}-0.008137 &\hspace*{-1ex}-0.014196
& \hspace*{-1ex}-0.019926 &\hspace*{-1ex}-0.02541
&\hspace*{-1ex}-0.03070\\
$C_2^{(\gamma_l)}$& 0.077 & 0.065533 & 0.056085 & 0.047668
& 0.039911 & 0.03265 & 0.02576 \\
$C_2^{(\varphi)}$&  --- & 0.023210 & 0.004723 &
\hspace*{-1ex}-0.011535&\hspace*{-1ex}-0.026221
&\hspace*{-1ex}-0.03965&\hspace*{-1ex}-0.05203\\
$C_{2}^{(\omega_l)}$&\hspace*{-1ex}-0.630  &\hspace*{-1ex}-0.482459 &
 \hspace*{-1ex}-0.361628 &\hspace*{-1ex}-0.253233 &
\hspace*{-1ex}-0.152736&\hspace*{-1ex}-0.05818 & 0.03190
\\\hline
\end{tabular}\medskip
\end{table}

In deriving the coefficients of the critical exponents $\alpha_l$,
$\beta_l$, and $\gamma_l$, we utilized the familiar hyperscaling
and scaling relations
\cite{HLS75,Sel92}
\begin{equation}\label{alphhsr}
\alpha_l=2-(d-m)\,\nu_{l2}-m\,\nu_{l4}\;,
\end{equation}
\begin{equation}\label{betahsr}
\beta_l={\nu_{l2}\over 2}\,(d-m-2+\eta_{l2})+{\nu_{l4}\over 2}\,m\;,
\end{equation}
and
\begin{equation}\label{gammasr}
\gamma_l=\nu_{l2}\,(2-\eta_{l2})=\nu_{l4}\,(4-\eta_{l4})\;,
\end{equation}
respectively. Specifically, the first one, Eq.~(\ref{alphhsr}), in
conjunction with Eqs.~(\ref{etataueps}) and (\ref{nul4eps}) yields
\begin{equation}
\alpha_l={4{-}n\over n{+}8}\,{\epsilon\over 2}+C_2^{(\alpha_l)}(n,m)\,\epsilon^2
+O(\epsilon^3)\;,
\end{equation}
with
\begin{equation}
C_2^{(\alpha_l)}(n,m)= {1\over 4}\,{n{+}2\over n{+}8}-{\left(4-{m\over  
2}\right)}\,C_2^{(\nu_{l2})}(n,m)-
m\,C_{2}^{(\nu_{l4})}(n,m)\;.
\end{equation}

Note that, to first order in  $\epsilon$, the expansions of the critical exponents
are independent of $m$. This means that one can set $m=0$. Hence
the expansions to first order in $\epsilon$
of all  critical exponents of the Lifshitz point that have well-defined analogs
for the usual isotropic ($m=0$) critical theory coincide with those of the latter,
which can be looked up in textbooks \cite{Ami84}. Examples
of such critical exponents
are $\nu_{l2}$, $\alpha_l$, $\beta_l$, and $\gamma_l$.

The source of this $m$ independence is the following.
The operator product expansion (OPE) of the theory considered here, for $\epsilon>0$,
is a straightforward extension of the familiar one of the isotropic ($m=0$)  
$\phi^4$ theory.  Proceeding by analogy with
 chapter 5.5 of Ref.~\cite{Car96b}), one can convince oneself that
the $O(\epsilon)$ corrections to the critical exponents are given by simple
ratios of OPE expansion coefficients. These do not require the explicit
computation of Feynman graphs but  follow essentially from combinatorics.
For critical exponents with an $m=0$ analog, this has the
above-mentioned consequence. The difference between the cases of a Lifshitz
point and of a critical point manifests itself in the $O(\epsilon)$ expressions
of these exponents only through the modified, $m$-dependent value of   
$\epsilon=4-d+m/2$, a difference that disappears for $m=0$.

\begin{figure}[htb] 
\includegraphics[width=\textwidth]{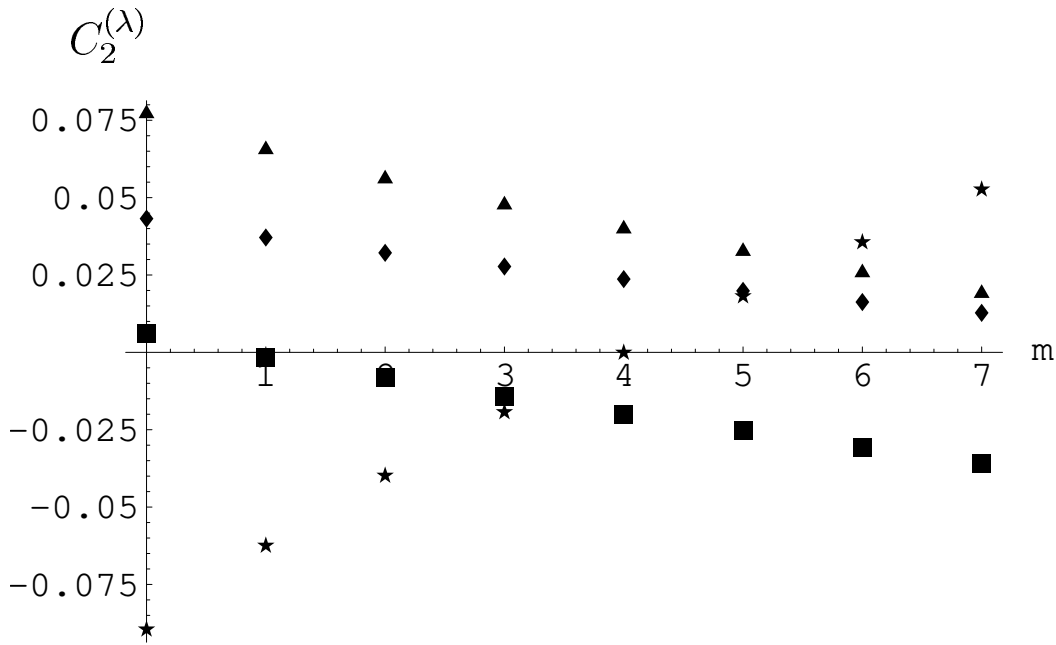}
\caption{Second-order coefficients $C_2^{(\lambda)}(n,m)$ of the exponents  
$\lambda=\nu_{l2}$ ($\blacklozenge$), $\alpha_l$ ($\bigstar$),
$\beta_l$  ($\blacksquare$), and $\gamma_l$ ($\blacktriangle$)
for $n=1$ and $m=0,\dots ,7$.}%
\label{fce2}%
\end{figure}

In Fig.~\ref{fce2} the coefficients $C_2^{(\lambda)}$
of the exponents $\lambda=\nu_{l2}$, $\alpha_l$, $\beta_l$, and $\gamma_l$
for the case $n=1$ are displayed as  functions of $m$. The results  indicate
that these coefficients depend in a smooth and monotonic fashion on $m$,
approaching the familiar isotropic $m=0$ values linearly in $m$.
Owing to the above-mentioned independence of the $O(\epsilon)$ expressions
of these critical exponents, this behavior carries over to the numerical estimates
one gets for the critical exponents in three dimensions by extrapolation of  
our $O(\epsilon^2)$ results. We shall see this explicitly shortly
(see Sec.~\ref{ssec:extra3D}). However, before turning to this matter, let us
briefly convince ourselves that our analytic two-loop series expressions
for those critical exponents, renormalization factors, etc.\ that
remain meaningful for $m= 0$ go over into the corresponding well-known results
of the usual isotropic $\phi^4$ theory for a critical point.

\subsection{The limit $m\to 0$}\label{ssec:mzerolim}

In this limit, $d$ `perpendicular' coordinates, but no `parallel' ones remain. Hence
 ${\boldsymbol{x}}$ can be identified with ${\boldsymbol{x}}_\perp$, and
the free propagator (\ref{Lsprop}) becomes
\begin{equation}\label{propmzero}
G({\boldsymbol{x}})= x^{-2+\epsilon}\,\Phi(\upsilon{=}0;m{=}0,d{=}4-\epsilon)=
\frac{1}{4}\,\pi^{-d/2}\,x^{-2+\epsilon}\;.
\end{equation}
The $\upsilon=0$ value of $\Phi(\upsilon)$ can be read off from the $k=0$ term of the
Taylor series (\ref{sinsum}). The expression on the far right of  
Eq.~(\ref{propmzero})
is indeed the familiar massless free propagator  $[-\triangle]  
^{-1}({\boldsymbol{x}})$.

We must now determine the $m\to 0$ limits of the integrals $j_\phi$, $j_\sigma$,
$j_\rho$, and $J_u$ (i.e., $j_u$), in terms of which our analytic results given in
Secs.~\ref{ssec:renfac}, \ref{ssec:betaf}, and \ref{ssec:epsexp}
are expressed. We assert that the correct limiting values
are
\begin{equation}\label{jmzeros}
j_\phi(0)=2\;,\quad j_\sigma(0)=j_\rho(0)=j_u(0)=0\;,\quad J_u(0)=\frac{1}{2}\;.
\end{equation}

To see this, note that $m$-dimensional integrals
$\int\!\d^m\upsilon\,f_m(\upsilon)=
S_m\,{\int_0^\infty}\!\d\upsilon\, f_m(\upsilon) \,\upsilon^{m-1}$ should approach
their zero-dimensional analog, namely $f_0(\upsilon{=}0)$, as $m\to 0$. In the case
of $j_\phi$, we have $f_0(0)=S_4\,\Phi^3(0;0,4)/F_{0,0}^{2}=2$. The
respective values $f_0(0)$ for $j_\sigma$ and $j_\rho$ vanish
because  of the explicit factors of $\upsilon^4$ and $\upsilon^2$ appearing in their
integrands. On the other hand, the vanishing of  $j_u(0)$ is due to the factor
$\Theta(\upsilon)$ of its integrand and the fact that $\Theta(0)=0$ according to
the Taylor series (\ref{Theta}). Finally, the value of $J_u(0)$ given above
follows from $j_u(0)=0$ via Eq.~(\ref{Ju}).

If one sets $m=0$ in the series expansions of  quantities whose analogs
retain their significance in the case of the usual critical-point theory,
e.g., in Eqs.~(\ref{Zphi}), (\ref{Ztau}), (\ref{Zu}),
(\ref{betau}), (\ref{ustar}), (\ref{etataueps}), and (\ref{omega}) for  
$Z_\phi$, $Z_\tau Z_\phi$, $Z_uZ_\phi^2Z_\sigma^{m/4}$, $\beta_u$, $u^*$,  
$\eta_\tau^*$,
and $\omega_{l2}$, utilizing the above values (\ref{jmzeros}) of the integrals,
one recovers the familiar two-loop results for the standard $\phi^4$ theory.

Let us also mention that (because of the factor $S_m\sim m$ in $B_m$)
the integrals $j_\sigma$, $j_\rho$, and $j_u$ vanish \emph{linearly}
in $m$ as $m\to 0$. Therefore, quantities like $Z_\sigma Z_\phi$,
$Z_\rho Z_\phi Z_\sigma^{1/2}$, $\eta_{l4}$, $\nu_{l4}$,
$\eta_\rho^*$, $\theta$, etc.\ that involve ratios such as $j_\sigma(m)/m$
\emph{have a finite $m\to 0$ limit.} It is conceivable that the $m\to 0$  
limits of these
quantities might turn out to have significance for appropriate problems.
However, we shall not further consider this issue here.

\subsection{The case of the isotropic Lifshitz point}\label{ssec:isoLP}

In the case of an \emph{isotropic Lifshitz point} one has $d=d_\|$ and $d_\perp=0$.
In a conventional $\epsilon$ expansion one would expand about  $d^*(8)=8$,
setting $d=d_\|=8-\epsilon_\|$. Results to order $\epsilon_\|^2$  that have been
obtained in this fashion for
the correlation exponents $\eta_{l2}$ and $\eta_{l4}$ and the correlation-length
exponents $\nu_{l2}$ and $\nu_{l4}$ can be found in Ref.\ \cite{HLS75b}.%
\footnote{We have checked these results by means of an independent calculation,
using dimensional regularization and minimal subtraction of poles.}

Since the constraint $d=d_\|$ implies that both $d$ and $d_\|$ vary as
$\epsilon_\|$ is varied, it may not be immediately clear that results for this case
can be extracted from our $\epsilon$ expansion at fixed $d_\|=m$.
Let us choose a fixed $m=8-\epsilon_\|$ and utilize the $\epsilon$ expansion.
This yields
\begin{equation}\label{nuexp}
\lambda(n,m,d)=\lambda^{(0)}+\lambda^{(1)}(n)\,\epsilon+\lambda^{(2)}(n,m)\,\epsilon^2+O(\epsilon^3)
\end{equation}
with $\epsilon=8-d-\epsilon_\|/2$, where $\lambda$ means any of the critical  
exponents
considered in Sec.~\ref{ssec:epsexp} that remain meaningful in the case of an  
isotropic
Lifshitz point, such as $\eta_{l4}$, $\nu_{l4}$, $\varphi$, $\alpha_l$,  
$\beta_q$, and $\omega_{l4}$. As indicated in Eq.~(\ref{nuexp}), the  
coefficients of the terms of orders $\epsilon^0$ and $\epsilon$ do \emph{not}  
depend on $m$. We  now set
$d=m$, which implies that $\epsilon=\epsilon_\|/2$. The upshot is the following:
In order to obtain from our $\epsilon$-expansion results the
dimensionality expansion of the critical
exponents of the isotropic Lifshitz point  about $d=8$ to second order,
we  must  simply replace the second-order coefficients $\lambda^{(2)}(n,m)$ by  
their limiting values $\nu^{(2)}(n,8^-)$ and identify $\epsilon$ with  
$\epsilon_\|/2=(8-d)/2$.

The limiting values of the integrals $j_\phi$, $j_\sigma$, $j_\rho$, and $J_u$
are
\begin{equation}\label{jmdvalues}
j_\phi(8^-)=0\;,\quad j_\sigma(8^-)=96\;,\quad j_\rho(8^-)=12\;,\quad  
J_u(8^-)=-\frac{1}{6}\;.
\end{equation}
To see this, note that the factor $B_m$ appearing in $j_\phi$,\ldots,  $j_u$ varies
$\sim (8{-}m)$ near $m=8$. In the case of $j_\phi$, the integral that multiplies
$B_m$ has a finite $m\to 8$ limit, so $j_\phi(8^-)$ vanishes. By contrast, the
corresponding integrals pertaining to $j_\sigma$ and $j_\rho$ have a pole of first
order at $m=8$. We have in $j_\sigma$,
\begin{eqnarray}
{\int_0^\infty}\!\d\upsilon\,\upsilon^{m+3}\,\Phi^3(\upsilon;m,d^*(m))&=&
{\int_0^\infty}\!\d
w\,w^{7-m}\,f(w)\,[1+O(8-d)]\nonumber\\&=&{f(0)\over8-m}+O{\left[(8{-}m)^0\right]}\;,
\end{eqnarray}
where $f(w)$ is the function $f(w)\equiv [w^{-4}\,\Phi(1/w;8,8)]^3$ whose value
$f(0)=(2\,\pi)^{-12}$ follows from Eq.~(\ref{asPhid*}). Using this together with
 ${B^\prime}(8)=-3 \times2^{17}\,\pi^{12}$ yields the above result for  
$j_\sigma(8^-)$.
The value of $j_\rho(8^-)$ follows in a completely analogous manner.

The computation of
$J_u(8^-)$ is somewhat more involved because the integral giving
$j_u/B_m$ has  poles of  \emph{second and first order.} The second-order pole,
due to the appearance of  the term $\sim\ln\upsilon$  in the large-$\upsilon$ form
(\ref{eq:asbehTheta}) of the scaling function $\Theta(\upsilon)$, produces a
first-order pole
in $j_u$, which cancels the pole resulting from the contribution  
$-\frac{1}{2}\,\psi(2{-}{m\over 4})$ to $J_u$ [cf.\ Eq.~(\ref{Ju})].
The first-order pole of the integral results in the
finite value (\ref{jmdvalues}) of $J_u(8^-)$.

Upon substituting the values (\ref{jmdvalues}) into the respective
 $\epsilon$-expansion results
of Sec.~\ref{ssec:epsexp} and setting $2\,\epsilon=\epsilon_\|=8-d$, we recover
indeed the results of Ref.~\cite{HLS75b} for the correlation  exponent  
$\eta_{l4}$ and the correlation-length exponent $\nu_{l4}$:
\begin{equation}
\eta_{l4}(m{=}d)=-{3\over 20}{n+2\over (n{+}8)^2}\,\epsilon_\|^2+O(\epsilon_\|^3)\;,
\quad\epsilon_\|\equiv 8-d\;,
\end{equation}
and
\begin{equation}
\nu_{l4}(m{=}d)={1\over 4}+{n+2\over 16(n{+}8)}\,\epsilon_\|
+{(n+2)(15n^2+89n+4)\over 960(n{+}8)^3}\,\epsilon_\|^2+O(\epsilon_\|^3)\;.
\end{equation}
Furthermore, we can infer the previously unknown series expansions of
the remaining exponents of the isotropic Lifshitz point.
Specifically for the wave-vector exponent, we find that
\begin{equation}\label{betaqiso}
\beta_q(m{=}d)=\frac{2+\eta_\sigma^*}{4(1+\eta_\rho^*)}=\frac{1}{2}+
{21\over 40}\,{n{+}2\over (n{+}8)^2}\,\epsilon_\|^2+O(\epsilon^3_8)\;.
\end{equation}
Another significant exponent  is the crossover exponent $\varphi$. Its $8-d$
expansion follows from Eq.~(\ref{betaqiso}) via the scaling law
$\varphi=\nu_{l4}/\beta_q$. The one of the  correction-to-scaling
exponent (\ref{omegal4def}) becomes
\begin{equation}
\omega_{l4}(m{=}d)=\epsilon_\|+ \frac{202 {+}  41\,n}{30\,( n {+} 8)^2}\,
\epsilon_\|^2 +O(\epsilon_\|^3)\;.
\end{equation}

\subsection{Series estimates of  the critical exponents for $d=3$ dimensions}
\label{ssec:extra3D}

We now wish to exploit our $\epsilon$-expansion results of the foregoing subsections
to obtain numerical values of the critical exponents in $d=3$ dimensions.
We shall mainly consider the cases of uniaxial Lifshitz points ($m=1$)
for order-parameter dimensions $n=1, 2, 3$ and of biaxial ($m=2$) Lifshitz  
points for $n=1$.
Of particular interest is the case $m=n=1$, which is realized by the Lifshitz point
of the ANNNI model \cite{Sel92,FS80} and is encountered in many experimental systems.

Cases with $m\ge 2$ are of limited interest whenever $n\ge 2$, for the  
following reason.
If a  Lifshitz point exists, then low-temperature spin-wave-type excitations whose
frequencies vary as
$\omega_{\boldsymbol{q}}=\sigma_0\,q_\|^4+q_\perp^2$ as $\boldsymbol{q}\to 0$
must occur. By analogy with the Mermin-Wagner theorem \cite{MW66} one concludes
that such excitations would destabilize an ordered phase in dimensions
$d\le d_*(m)=2+m/2$, ruling out the possibility of a
spontaneous breaking of the $O(n)$ symmetry at temperatures $T>0$ for
such values of $d$. (This conclusion is in complete accordance
with Grest and Sak's work \cite{GS78} based on nonlinear sigma models.)
Hence in three dimensions one is left with the case $m=1$ of a uniaxial  
Lifshitz point
if $n\ge 2$.

In Table \ref{tab:te} we list numerical estimates of the critical exponents
for $d=3$, $n=1$, and $m=1,2,\ldots,6$. For comparison, we also included
the $m=0$ values of those critical and correction-to-scaling exponents
that go over into their standard counterparts $\nu$, $\gamma$, $\alpha$,
$\beta$, and $\omega$ for a critical point.%
\footnote{Numerical estimates of the correlation exponents
$\eta_{l2}$ and $\eta_{l4}$ are not included in Table \ref{tab:te},
as they can be found in I.}
As is explained in the caption, these estimates were either obtained by
setting $\epsilon=d^*(m)-3$ in the $O(\epsilon^2)$ expressions of the exponents
or else via $[1/1]$ Pad\'e approximants.

\begin{table}[htb]
\caption{Numerical estimates for the critical exponents with $n=1$ and $d=3$. The
values marked by superscripts $(\epsilon)$ were obtained by setting $\epsilon=1+m/2$
in the expansions to order $\epsilon^2$ of the exponents; those marked by
superscripts $[1/1]$ were determined from $[1/1]$ Pad\'e approximants whose  
parameters were  fixed by the requirement that the respective expansions
to second order in $\epsilon$ are reproduced.
 \label{tab:te}}
\begin{tabular}{lccccccc}\hline
$m$&0&1&2&3&4&5&6\\\hline
$\nu_{l2}^{(\epsilon)}$ & 0.627 & 0.709 & 0.795 & 0.882 & 0.963
                           & 1.035 & 1.093 \\
$\nu_{l2}^{[1/1]}$ &0.673 & 0.877 & 1.230 & 1.742 & 2.19\phantom{0}
                           & 2.26\phantom{0} & 2.02\phantom{0} \\
$\nu_{l4}^{(\epsilon)}$ &---  & 0.348 & 0.387 & 0.423 & 0.456
                           & 0.482 & 0.500 \\
$\nu_{l4}^{[1/1]}$&--- & 0.396 & 0.482 & 0.561 & 0.606
                           & 0.609 & 0.585 \\
$\gamma_l^{(\epsilon)}$& 1.244 & 1.397 & 1.558 &
 1.715 & 1.859 & 1.983 & 2.08\phantom{0} \\
$\gamma_l^{[1/1]}$&1.310 & 1.609 & 2.02\phantom{0}
& 2.46\phantom{0} &2.78\phantom{0} &
 2.86\phantom{0} & 2.75\phantom{0} \\
$\alpha_l^{(\epsilon)}$ & 0.077 & 0.110 & 0.174 & 0.296 & 0.499
                           & 0.806 & 1.24\phantom{0} \\
$\alpha_l^{[1/1]}$& 0.108 & 0.160 & 0.226 & 0.323 & 0.499
                           & 0.94 & 4.6\phantom{00} \\
$\beta_l^{(\epsilon)}$ & 0.340 & 0.247 & 0.134 &\hspace*{-1ex}-0.005 &
\hspace*{-1ex}-0.18\phantom{0}&\hspace*{-1ex}-0.39\phantom{0} &
\hspace*{-1ex}-0.7\phantom{00} \\
$\beta_l^{[1/1]}$& 0.339 & 0.246 & 0.131 &\hspace*{-1ex}-0.029  
&\hspace*{-1ex}-0.28\phantom{0}
                           &\hspace*{-1ex}-0.75\phantom{0}  
&\hspace*{-1ex}-2.0\phantom{00} \\
$\varphi^{(\epsilon)}$ & ---& 0.677 & 0.686 & 0.636 & 0.514
                           & 0.306 & 0.001 \\
$\varphi^{[1/1]}$&--- & 0.715 & 0.688 & 0.654 & 0.628
                           & 0.609 & 0.595 \\
$\omega_{l2}^{(\epsilon)}$ & 0.370 & 0.414 & 0.553 & 0.240 &
 \hspace*{-1ex}-0.255  &\hspace*{-1ex}-0.930 &\hspace*{-1ex}-1.786 \\
$\omega_{l2}^{[1/1]}$& 0.614 & 0.870 & 1.161 & 1.313 & 1.439
                           & 1.545 & 1.635\\\hline
\end{tabular}
\end{table}

According to Table \ref{tab:C2n=1}, the coefficients of most of these series  
with $n=1$
do \emph{not\/} alternate in sign.
Exceptions are the ones of $\alpha_l$ and $\omega_{l2}$ for small values of $m$,
that of $\beta_l$ for $m=0$ (i.e., of the usual critical index $\beta$), and
the one of $\varphi$ for larger values of $m$.
For $d=3$, the second-order contributions grow very rapidly
as $m$ increases because of the factor $\epsilon^2=(1+m/2)^2$.
Therefore the numerical estimates become less reliable for large $m$.
This effect is more pronounced for $[1/1]$ estimates from non-alternating series
than for the corresponding direct evaluations at $d=3$
(marked by superscripts ${(\epsilon)}$). The better-behaved expansions
yield smaller differences between these two kinds of estimates.
In unfavorable cases  with rather large $\epsilon$ we reject the $[1/1]$
estimates for the non-alternating series, which tend to overestimate the
values of the corresponding exponents. Instead we prefer the direct evaluations
at $d=3$.

A reversed situation occurs for $\alpha_l$ and $\omega_{l2}$
with $m=0, 1,  2$. The respective series are alternating; they have
negative $O(\epsilon^2)$ corrections, which tend to
underestimate the values of the exponents for $d=3$ in direct evaluations of
the $O(\epsilon^2)$ expressions.
On the other hand, the $[1/1]$ approximants for these series%
\footnote{The $\epsilon$ expansions
of $\alpha_l$ and $\omega_{l2}$ start at order $\epsilon$.
We add unity to these series, construct
the $[1/1]$ approximants, and subsequently subtract unity from the
resulting numerical values of the approximants.}
seem to do a better job, suppressing the influence of the second-order
corrections in a correct way.
We believe that $\alpha_l^{[1/1]}(m{=}n{=}1)\simeq 0.160$ belongs
to our best numerical $d=3$ estimates that are obtainable
from the individual $\epsilon$ expansions.

In the case of $\omega_{l2}$, the second-order correction is much larger.
While therefore less accurate numerical estimates must be expected, the structure
of the $\epsilon$ expansion for $\omega_{l2}$ suggests nevertheless that
this correction-to-scaling exponent  should have a larger value  than its
$m=0$ counterpart  $\omega$ for  the critical point.  (Recall that the latter  
has a value
 close to $0.8$ \cite{GZJ98}). Our best estimate is the $[1/1]$
value
\begin{equation}\label{omegal2est}
\omega_{l2}\simeq 0.870\;.
\end{equation}

In order to obtain improved estimates we proceed as follows. We
choose the ``best'' $d=3$ estimates we can get  from the apparently best-behaved
$\epsilon$ expansions of certain exponents, express the remaining critical indices
in terms of the former, and compute their implied values.
Thus we select from Table \ref{tab:te} the numbers
\begin{equation}\label{benul4}
\nu_{l4}^{(\epsilon)}(d{=}3;m{=}1,n{=}1)\simeq 0.348 \quad\mbox{and}\quad
\alpha_l^{[1/1]}(3;1,1)\simeq 0.160\;,
\end{equation}
which we complement  by our estimate
\begin{equation}\label{beetal4}
\eta_{l4}(3;,1,1)\simeq -0.019
\end{equation}
from I. Substituting these into the second one of the scaling relations
(\ref{gammasr})
for $\gamma_l$ and the hyperscaling relation (\ref{alphhsr}) for $\alpha_{l}$
yields
\begin{equation}\label{begammal}
\gamma_l(3;1,1) \simeq 1.399\quad\mbox{and}\quad
\nu_{l2}(3;1,1) \simeq  0.746\;,
\end{equation}
respectively,
from which in turn the values
\begin{equation}\label{benul2}
\eta_{l2}(3;1,1) \simeq 0.124\quad\mbox{and}\quad
\beta_l(3;1,1) \simeq 0.220
\end{equation}
follow via the scaling relations $\eta_{l2}=2-\gamma_l/\nu_{l2}$ and  
$\beta_l=(2-\alpha_l-\gamma_l)/2$.

Likewise, the choices
\begin{equation}\label{bevarphinul4}
\varphi(3;1,1)\simeq\varphi^{(\epsilon)}= 0.677
\quad\mbox{and}\quad
\nu_{l4}(3;1,1)\simeq \nu_{l4}^{(\epsilon)}=0.348
\end{equation}
give for the wave-vector exponent%
\footnote{Note that  Eq.~(77) of I, which recalls the conventional
definition of  $\beta_q$,
contains a misprint: the variable $\tau$ should be replaced by $q$.}
\begin{equation}
\beta_q={\nu_{l4}/ \varphi}
\end{equation}
the estimate
\begin{equation}\label{bebetaq}
\beta_q(3;1,1)\simeq 0.514\;,
\end{equation}
which is fairly close to the value $\beta_q\simeq 0.519$ of I.
We consider the  values  (\ref{omegal2est})--(\ref{bevarphinul4}) and (\ref{bebetaq})
as our best estimates for these 10 critical exponents.

\begin{table}[htb]
\caption{Critical exponents for $m=1$, $n=1$, $2$, $3$, and $d=3$. The row  
marked MF gives the mean-field values; rows  marked $O(\epsilon)$ and  
$O(\epsilon^2)$ list the values obtained by setting $\epsilon=3/2$ in the  
expansions to first and
second order in $\epsilon$, respectively.  The remaining rows contain the estimates
from  $[1/1]$ Pad\'e approximants and our ``best'' estimates (`Scal.') obtained
via scaling relations in the manner explained in the main text.
\label{tres}}\
\begin{tabular}{c|c|cccc|cc|cc}\hline
&&\multicolumn{4}{c|}{$n=1$}&\multicolumn{2}{c|}{$n=2$}&\multicolumn{2}{c}{$n=3$}\\
&MF&$O(\epsilon)$&$O(\epsilon^2)$&$[1/1]$&Scal.&
$O(\epsilon)$&$O(\epsilon^2)$&$O(\epsilon)$&$O(\epsilon^2)$\\\hline
$\nu_{l2}$&${1\over 2}$& 0.625& 0.709& 0.877& 0.746 & 0.65\phantom{0} & 0.757&  
0.67\phantom{0}&0.798\\
$\nu_{l4}$&${1\over 4}$&0.313&0.348&0.396& 0.348& 0.325& 0.372& 0.335& 0.392 \\
$\alpha_l$ &0&0.25\phantom{0}&0.110&0.160&0.160&
0.15\phantom{0}&\hspace*{-1ex}-0.047&0.068&\hspace*{-1ex}-0.178\\
 $\beta_l$ &${1\over 2}$&0.25\phantom{0}&0.247& 0.246&0.220&0.275&0.276&0.295&0.301\\
 $\gamma_l$ &1&1.25\phantom{0}&1.397&  
1.609&1.399&1.3\phantom{00}&1.495&1.34\phantom{0}&1.576\\
 $\varphi$ &${1\over  
2}$&0.625&0.677&0.715&0.677&0.65\phantom{0}&0.725&0.67\phantom{0}&0.765\\
 $\beta_q$&${1\over  
2}$&0.5\phantom{00}&0.519&&0.514&0.5\phantom{00}&0.521&0.5\phantom{00}
&0.521\\
$\eta_{l2}$ &0&0&0.039&&0.124&0&0.042&0&0.044\\
$\eta_{l4}$&0&0&\hspace*{-1ex}-0.019&&  
\hspace*{-1ex}-0.019&0&\hspace*{-1ex}-0.020&0&\hspace*{-1ex}-0.021\\
$\omega_{l2}$&0&1.5\phantom{00}&0.414&0.870&&1.5\phantom{00}
&0.466&1.5\phantom{00}&0.517\\\hline
\end{tabular}
\end{table}

Table \ref{tres}  presents an overview of our numerical findings. For convenience,
the mean-field results are included along with the values
the $\epsilon$ expansions to first and second order take at $\epsilon=3/2$.
The case  $m=1$ with $n=2$ corresponds to the Lifshitz point of the
axial next-nearest-neighbor XY (ANNNXY) model, which Selke
studied many years ago by means of Monte Carlo simulations \cite{Sel80}.

In Table \ref{tab:otherres} we have gathered
the available experimental results for critical exponents together with
estimates obtained from Monte Carlo calculations and high-temperature series
analyses. As one sees, our field-theory estimates are
in a good agreement with the Monte Carlo results. The experimental value for
$\alpha_l(3;1,1)$ deviates appreciably both from all theoretical estimates (including
ours) as well as from the Monte Carlo results,
 and is probably not very accurate. On the other hand, the very good agreement
of our field-theory estimates with the most recent Monte Carlo estimates
by Pleimling and Henkel \cite{PH01}  (which we expect to be the most accurate ones)
is quite encouraging. Certainly, renewed experimental efforts for determining
the values of the critical exponents in a more complete and more precise way
would be most welcome.

\begin{table}[htb]
\caption{Values of critical exponents for uniaxial Lifshitz points ($m{=}1$).
`Exp'  means summarized experimental results, taken from Ref.~\cite{ZSK00};
`HT'  denotes the high-temperature series estimates of Ref.~\cite{MF91}.
The rest are the Monte Carlo results of  Refs.~\cite{Sel78} (MC1),   
\cite{KS85} (MC2),
\cite{PH01} (MC3), and \cite{Sel80} (MC), respectively.
\label{tab:otherres}}
\begin{tabular}{@{$\;$}l@{$\;$}|cccccc@{$\;$}}\hline
&$\nu_{l4}$&$\alpha_l$&$\beta_l$&$\gamma_l$&$\varphi$&$\beta_q$\\\hline
&\multicolumn{6}{c}{$n=1$}\\\hline
Exp&  & 0.4--0.5 &  &  &0.60--0.64 &0.44--0.49\\
HT & $0.41{\pm}0.03$ & 0.20${\pm}$0.15 &  & 1.62${\pm}$0.12 &  &0.5\\
MC1 &  &  & 0.21${\pm}$0.03 & 1.36${\pm}$0.005 &  &\\
MC2 & 0.33${\pm}$0.03 & 0.2 & 0.19${\pm}$0.02 & 1.4${\pm}$0.06 &  &\\
MC3 &  & 0.18${\pm}$0.03 & 0.235${\pm}$0.005 & 1.36${\pm}$0.03 &  &\\\hline
&\multicolumn{6}{c}{$n=2$}\\\hline
MC  &  & 0.1${\pm}$0.14 & 0.20${\pm}$0.02 & 1.5${\pm}$0.1 &  &\\\hline
\end{tabular}

\end{table}

\section{Concluding remarks}\label{sec:Concl}

The field-theory models (\ref{Ham}) were introduced more than 25 years ago
to describe the universal critical behavior at $m$-axial Lifshitz points  
\cite{HLS75b}.
While some field-theoretic studies based on the $\epsilon$ expansion about
the upper critical dimension $d^*(m)$ emerged soon afterwards,
 these were limited to first order in $\epsilon$, or restricted to special values
of $m$ or to a subset of critical exponents, or challenged by discrepant results
(see the references cited in the introduction).
Two-loop calculations for general values of $m$ appeared to be hardly feasible
because of the severe calculational difficulties that must be overcome.

Complementing our previous work in I, we have presented here a full
two-loop RG calculation for the models (\ref{Ham}) in $d=d^*(m)-\epsilon$
dimensions, for general values of $m\in(0,8)$. This enabled us to
compute the $\epsilon$ expansions of all critical indices of the considered
$m$-axial Lifshitz points to second order in $\epsilon$.
We employed  these results in turn to determine field-theory estimates for the values
of these  critical exponents in three dimensions. Although the accuracy of
these estimates clearly is not competitive with the impressive
precision that has been achieved by the best field-theory estimates
for critical exponents of conventional critical points
(based on perturbation expansions to much higher orders and powerful
resummation techniques \cite{GZJ98,ZJ96}), they are in very good agreement
with recent Monte Carlo results for the uniaxial scalar case $m=n=1$.
We hope that our present work will stimulate new efforts,
both by experimentalists and theorists, to
investigate the critical behavior at Lifshitz points.

There is a number of promising directions in which our work could be extended.
For example, building on it, one could compute other universal quantities, such as
amplitude ratios and scaling functions, via the $\epsilon$ expansion.

A particular interesting and challenging question is whether
the generalized invariance found by Henkel \cite{Hen97}
for systems whose  anisotropy exponents take the
rational values $\theta=2/\wp$,
$\wp=1,2,3\ldots$ can be generalized to other, irrational values.
That such an extension exists, is not at all clear since
the condition $\theta=2/\wp$ is utilized in Henkel's work to ensure
that the algebra closes. But if such an extension can be found, then
the invariance under this larger group of transformations should manifest
itself through properties of  the theories' scaling functions in $d<d^*(m)$
dimensions, which could be checked by means of the $\epsilon$ expansion.
Furthermore, even if  an extension cannot be found,
one should be able to benefit from the invariance properties
of the free theory (with $ \theta=1/2$) when computing
the $\epsilon$ expansion of anomalous dimensions of  composite operators
in a similar extensive fashion
as in the case of the standard critical-point $\phi^4$ theory  
\cite{KWP93,KW94,Keh95}.

An important issue awaiting clarification arises when $m\ge 2$.
In the class of models (\ref{Ham}) studied here, the quadratic
fourth-order derivative term
was taken to be isotropic in the subspace ${\mathbb{R}}^m$.
However, in general further fourth-order derivatives
cannot be excluded. That is, the term  
$(\sigma_0/2)\,(\triangle_\|\boldsymbol{\phi})^2$
should be generalized to
\begin{equation}
(\sigma_0/2)\, w_a\, T^{(a)}_{ijkl}\,(\partial_i\partial_j{\boldsymbol{\phi}})
\partial_k\partial_l\,{\boldsymbol{\phi}}\;,
\end{equation}
where the summation over $a$ comprises all totally symmetric fourth-rank tensors
$T^{(a)}_{ijkl}$ compatible with the symmetry of the considered
microscopic (or mesoscopic) model. The isotropic fourth-order derivative
term corresponds to $T^{(a{=}1)}_{ijkl}\equiv (\delta_{ij}\,\delta_{kl}
+\delta_{ik}\,\delta_{jl}
+\delta_{il}\,\delta_{jk})/3$ with $w_{1}\equiv 1$.

In order to give
a simple example of a system involving a further
quadratic fourth-order derivative term,
let us consider an $m$-axial  modification of the familiar uniaxial ANNNI   
model \cite{FS80,Sel92} that has
competing nearest-neighbor (nn) and next-nearest-neighbor interactions along
$m$ equivalent of the $d$  hypercubic axes (and only the usual nn bonds along the
remaining $d-m$ ones).
Owing to the Hamiltonian's hypercubic (rather than isotropic) symmetry in the  
$m$-dimensional subspace, just one other fourth-order derivative term,
corresponding to the tensor $T^{(2)}_{ijkl}=
\delta_{ij}\delta_{kl}\delta_{li}$, must generically occur
besides the isotropic one, in a coarse-grained description.
The associated interaction constant $w_2$ is dimensionless and hence
marginal at the Gaussian fixed point. To find out whether
the nontrivial ($u^*> 0$, $w_2=0$) fixed point considered throughout this work  
remains
infrared stable, one must compute the anomalous dimension
of the additional scaling operator that can be formed from
the above two fourth-order derivative terms.
This issue will be taken up in a forthcoming joint paper with R.~K.~P.\ Zia
\cite{DSZ01}, where we shall show that the associated crossover exponent,
to order $\epsilon^2$, is indeed positive.  Hence, deviations from $w_2=0$
correspond to a \emph{relevant perturbation} at the $w_2=0$, $u^*>0$ fixed  
point, which should destabilize it unless $m=1$.

Finally, let us mention that the potential of the $\epsilon$-expansion results
presented in this paper certainly has not  fully been exploited here.
When estimating the values of the critical exponents for $d=3$ dimensions,
we utilized only their $\epsilon$ expansions for a fixed \emph{integer}
number of the parameter $m$. However, our results hold also for \emph{noninteger}
values of  $m$. Making use of this fact, one should be able to extrapolate
to points  $(d=3,m=\mbox{integer})$ of interest in a more flexible
fashion, starting from any point on the critical curve $d=d^*(m)$ and
going along directions not perpendicular to the $m$ axis. By exploiting
this flexibility one should be able to improve the accuracy of the estimates.

Last but not least, let us briefly mention where the interesting reader can
find information about experimental results. Earlier experimental work
is discussed in Hornreich's and Selke's review articles \cite{Hor80,Sel92}.
A more recent summary of experimental results for the critical exponents
and other universal quantities of the Lifshitz point in MnP and
Mn$_{0.9}$Co$_{0.1}$P
can be found in Ref.~\cite{ZSK00} and its references.
(These results were partly quoted in Table \ref{tab:otherres}.)
However, the variety of experimental systems having
(or believed to have) Lifshitz points is very rich, ranging from
ferromagnetic and ferroelectric systems to polymer mixtures.
A complete survey of  the published experimental results on Lifshitz points
is beyond the scope of the present article.

\begin{ack}
It is our  pleasure to thank Malte  Henkel and Michel Pleimling for
informing us about their work \cite{PH01} prior to publication as well as
for discussions and correspondence.
We gratefully acknowledge support by the Deutsche Forschungsgemeinschaft (DFG)
via  the Leibniz program Di 378/2-1.
\end{ack}

\appendix
\section{Series representation and asymptotic expansion of $\Phi(\upsilon)$}  
\label{scf}
>From the integral representation (\ref{F}) of the scaling
function $\Phi(\upsilon)$ we find
\begin{eqnarray}\label{irepPhi}
\Phi(\upsilon)&=&\int_{{\boldsymbol{q}}_\|} \int_{{\boldsymbol{q}}_\perp}
{\e^{i{\boldsymbol{q}}_\perp\cdot{\boldsymbol{e}}_\perp}\,
\e^{i{\boldsymbol{q}}_\|\cdot \boldsymbol{\upsilon}}\over q_\perp^2+q_\|^4}
={\upsilon}^{-4+2\epsilon}\int_{{\boldsymbol{q}}_\|} \int_{{\boldsymbol{q}}_\perp}
{\e^{i\upsilon^{-2}{\boldsymbol{q}}_\perp\cdot{\boldsymbol{e}}_\perp}\,
\e^{i{\boldsymbol{q}}_\|\cdot{\boldsymbol{e}}_\|}\over q_\perp^2+q_\|^4}\;,
\end{eqnarray}
where ${\boldsymbol{e}}_\|=\boldsymbol{\upsilon}/\upsilon$ is an arbitrary
unit $m$-vector. The second form follows via
rescaling of the momenta; it lends itself for studying the
\emph{large}-$\upsilon$ behavior. The first one is appropriate for
deriving the small-$\upsilon$ expansion.

Upon utilizing the Schwinger representation
\begin{equation}\label{Schwinger}
\frac{1}{q_\perp^2+q_\|^4}={\int_0^\infty}\!{\d}s\,\e^{- s\,\big(  
q_\perp^2+q_\|^4\big)}
\end{equation}
for the momentum-space propagator in Eq.\ (\ref{irepPhi}),
we can perform the integration over
${\boldsymbol{q}}_\perp$ to obtain
\begin{eqnarray}\label{Phi2nd}
\Phi(\upsilon)&=&(4\,\pi)^{-d_\perp/2}\,{\upsilon}^{-4+2\epsilon}{\int_0^\infty}\!\d  
s\,s^{-d_\perp/2}\int_{{\boldsymbol{q}}_\|}\e^{-s\,q_\|^4 -
({4\, s\, \upsilon^4)}^{-1}}
\e^{i{\boldsymbol{q}}_\|\cdot{\boldsymbol{e}}_\|}\;.
\end{eqnarray}
Doing the angular integrations gives
\begin{equation}
\int_{{\boldsymbol{q}}_\|}\e^{-s\,q_\|^4}
\e^{i{\boldsymbol{q}}_\|\cdot{\boldsymbol{e}}_\|}=
(2\,\pi)^{-m/2}\,{\int_0^\infty}\!\d q\,q^{m/2}\,J_{m-2\over 2}(q)
\,\e^{-s\,q^4}\;.
\end{equation}
We insert this into Eq.~(\ref{Phi2nd}), expand the exponential
$\e^{-1/(4\,s\,\upsilon^4)}$ in powers of $\upsilon^{-4}$, and
integrate the resulting series termwise over $s$. This yields
\begin{eqnarray}\label{asePhi}
\Phi(\upsilon)&=&2^{-2-m}\,\pi^{-{d-1\over 2}}\,{\left({\upsilon\over  
2}\right)}^{2\,\epsilon-4}\sum_{k=0}^\infty
\frac{(-1)^k}{k!}\,\Phi_k\,{\left({\upsilon\over 2}\right)}^{-4k}
\end{eqnarray}
with
\begin{eqnarray}\label{Phik}
\Phi_k&=&{\Gamma{\left(\frac{m}{4}-1-k-\frac{\epsilon}{2}\right)}
\over 2^{6 -m+ 6k - 3\,\epsilon }}\;
 {\int_0^\infty}\!\d q\, {q^{4(1+k) - {\frac{m}{2}} - 2\,\epsilon }}\,
J_{m-2\over 2}(q)\nonumber\\&=&
{\frac{2\,\Gamma(2 + 2k - \epsilon )}
   {\Gamma{\left(\frac{m}{4}-\frac{1}{2}+\frac{\epsilon}{2}-k\right)}}}\;,
\end{eqnarray}
which is the asymptotic expansion (\ref{asexpPhihat}). The Taylor series
(\ref{sinsum}) can be derived along similar lines, starting from the
first form of the integral representation (\ref{irepPhi}).

\section{Laurent expansion of required vertex functions}\label{Lexp}

In this appendix we gather our results on the Laurent expansions of
those vertex functions whose pole terms determine the required renormalization
factors. It is understood that $\tau_0$ and $\rho_0$ are set to their critical
values $\tau_c=\rho_c=0$. For notational simplicity, we
introduce the dimensionless bare coupling constant
\begin{equation}
\check{u}_0\equiv\mu^{-\epsilon}\,\sigma_0^{-m/4}\,u_0=Z_u\,u
\end{equation}
and specialize to the
scalar case $n=1$. The generalization to the $n$-component case
involves the usual tensorial factors and contractions of the standard
$|\boldsymbol{\phi}|^4$ theory and should be obvious.

We use the notation $\tilde{\Gamma}(\{{\boldsymbol{q}}_j\})$
for the Fourier transforms
of vertex functions $\Gamma(\{\boldsymbol{x}_j\})$
(with the momentum-conserving $\delta$ function taken out):
\begin{equation}
\Gamma(\{\boldsymbol{x}_j\})=\int_{{\boldsymbol{q}}_1,\ldots,{\boldsymbol{q}}_N}
\tilde{\Gamma}(\{{\boldsymbol{q}}_j\})
\,(2\,\pi)^d\,\delta{\bigg(\sum_j
\boldsymbol{q}_j\bigg)}\,
\e^{i\sum_{j=1}^N\boldsymbol{q}_j\cdot
\boldsymbol{x}_j}\;.
\end{equation}

\subsection{Two-point vertex functions $\Gamma^{(2)}$ and %
$\Gamma^{(2)}_{(\napar\phi)^2}$}

>From our results obtained in I we find%
\footnote{We suppress diagrams involving the one-loop (sub)graph
$\,{\begin{texdraw}\drawdim pt \setunitscale 2.5   \linewd 0.3
\lellip rx:2.6 ry:2.8
\move(-3 -2.8) \rlvec(6 0)
\end{texdraw}}\,$
since the latter vanishes for $\tau=0$ if dimensional regularization
is employed, as we do throughout this paper.}
 \begin{equation}
\tilde{\Gamma}^{(2)}({\boldsymbol{q}})=\sigma_0 q_\|^4+q_\perp^2-
\raisebox{-4.25pt}{
\begin{texdraw}
\drawdim pt \setunitscale 2.5   \linewd 0.3
\lellip rx:5 ry:2.8
\move(-8 0)
\lvec(8 0)
\end{texdraw}}\;+O(u_0^3)
\end{equation}
with
\begin{equation}
\raisebox{-4.25pt}{
\begin{texdraw}
\drawdim pt \setunitscale 2.5   \linewd 0.3
\lellip rx:5 ry:2.8
\move(-8 0)
\lvec(8 0)
\end{texdraw}}\;
={\check{u}_0^2\over 6\,\epsilon}\,{\left[
{j_{\sigma}(m)\,\sigma_0\,q_\|^4\over 16\,m(m{+}2)}-
 \frac{{j}_{\phi}(m)\,q_\perp^2}{2\,(8-m)}\right]}+O(\epsilon^0)
\end{equation}
and
\begin{equation}
\tilde{\Gamma}^{(2)}_{(\napar\phi)^2}({\boldsymbol{q}},{\boldsymbol{Q}}=
{\boldsymbol{0}})=
q_\|^2-
\raisebox{-4.25pt}{
\begin{texdraw}
\drawdim pt \setunitscale 2.5   \linewd 0.3
\lellip rx:5 ry:2.8
\move(-8 0)
\lvec(8 0)
\move(0 2.8)
\fcir f:0 r:0.7
\rmove(-1.15 -1)\rlvec(0 2)
\rmove(2.25 -2)\rlvec(0 2)
\end{texdraw}}
\;+O(u_0^3)
\end{equation}
with
\begin{equation}
\raisebox{-4.25pt}{
\begin{texdraw}
\drawdim pt \setunitscale 2.5   \linewd 0.3
\lellip rx:5 ry:2.8
\move(-8 0)
\lvec(8 0)
\move(0 2.8)
\fcir f:0 r:0.7
\rmove(-1.15 -1)\rlvec(0 2)
\rmove(2.25 -2)\rlvec(0 2)
\end{texdraw}}=
{\check{u}_0^2\over 2\,\epsilon}\,
{j_{\rho}(m)\over 4\,m}\,q_\|^2+O(\epsilon^0)\;,
\end{equation}
where $\boldsymbol{Q}=\boldsymbol{0}$ is the momentum of the inserted operator  
$(\napar\phi)^2/2$; i.e., the insertion considered
is ${\int}\d^dx(\napar\phi)^2/2$.

\subsection{Four-point vertex function $\Gamma^{(4)}$}
The four-point vertex function was computed only to one-loop order
in I. To the order of two loops it reads%
\footnotemark[\value{footnote}]
\begin{eqnarray}\label{Gamma4}
\tilde{\Gamma}^{(4)}({\boldsymbol{q}}_1,\ldots,{\boldsymbol{q}}_4)&=&
u_0 -
\left(\;
\raisebox{-2.2pt}{\begin{texdraw}
\drawdim pt \setunitscale 2.5   \linewd 0.3
\lellip rx:4 ry:1.8
\move(4 0)\rlvec(2 2)
\move(4 0)\rlvec(2 -2)
\move(-4 0)\rlvec(-2 -2)
\move(-4 0)\rlvec(-2 2)
\end{texdraw}}+2\,\mathcal{P\/}\right)-
\left(\;
\raisebox{-2.2pt}{\begin{texdraw}
\drawdim pt \setunitscale 2.5   \linewd 0.3
\lellip rx:4 ry:1.8
\move(-4 0)\rlvec(-2 -2)
\move(-4 0)\rlvec(-2 2)
\move(8 0)
\lellip rx:4 ry:1.8
\move(12 0)\rlvec(2 2)
\move(12 0)\rlvec(2 -2)
\end{texdraw}}+2\,\mathcal{P\/}\right)
\nonumber \\[1em]&&\mbox{}
-
\left(\;
\raisebox{-8pt}{\begin{texdraw}
\drawdim pt \setunitscale 2.5   \linewd 0.3
\move(-7 -1.5)\rlvec(15 6)
\move(-7 1.5)\rlvec(15 -6)
\move(5 0)
\lellip rx:1 ry:3
\end{texdraw}}
+5\,\mathcal{P}\right)
+O(u_0^4)\\[1em]
&=&u_0{\left\{1-\sum_{(ij)=(12),(23),(24)}{\left[{\check{u}_0\over  
2}\,I_2(\check{\boldsymbol{q}}_{ij})-
{\check{u}^2_0\over 4}\,I_2^2(\check{\boldsymbol{q}}_{ij})
\right]}\right.}\nonumber\\
&&{\left.\mbox{}+\frac{\check{u}_0^2}{2}\,
{\left[I_4(\check{\boldsymbol{q}}_{12},\check{\boldsymbol{q}}_3)+5\,\mathcal{P}\right]}
\right\}}\;.
\end{eqnarray}
Here $\mathcal{P}$ means permutations (of the external legs). The hatted
momenta are dimensionless ones defined via
\begin{equation}\label{hatq}
\check{\boldsymbol{q}}=
(\check{\boldsymbol{q}}_\|,\check{\boldsymbol{q}}_\perp)\equiv
{\left(\sigma_0^{1/4}\,\mu^{-1/2}\,{\boldsymbol{q}}_\|,
\mu^{-1}\,{\boldsymbol{q}}_\perp\right)}\;,
\end{equation}
and
$\check{\boldsymbol{q}}_{ij}\equiv\check{\boldsymbol{q}}_{i}+\check{\boldsymbol{q}}_{j}$.
The integrals $I_2$ and $I_4$ are given by
\begin{equation}\label{I2def}
I_2({\boldsymbol{Q}})\equiv\int_{\boldsymbol{q}}\,
{1\over
q_\perp^2+q_\|^4}\,{1\over {\Big({\boldsymbol{q}}+{\boldsymbol{Q}}\Big)}_\perp^2
+{\left({\boldsymbol{q}}+{\boldsymbol{Q}}\right)}_\|^4
}
\end{equation}
and
\begin{equation}\label{I4def}
I_4({\boldsymbol{Q}},{\boldsymbol{K}})\equiv
\int_{{\boldsymbol{q}}}\,
{1\over
q_\perp^2+q_\|^4}\,{1\over {\left({\boldsymbol{q}}+{\boldsymbol{Q}}\right)}_\perp^2
+{\left({\boldsymbol{q}}+{\boldsymbol{Q}}\right)}_\|^4
}\,
I_2({\boldsymbol{q}}-{\boldsymbol{K}})\;.
\end{equation}

The pole term of $I_2({\boldsymbol{Q}})$ can be read off from
Eqs.\ (24) and (89) of I. However, in our two-loop calculation $I_2$
also occurs as a divergent subintegral. To check that the associated pole
terms are canceled by contributions involving one-loop counterterms, we
also need the finite part of $I_2$. The calculation simplifies considerably if
the momentum $\boldsymbol{Q}$ is chosen to have a perpendicular component
only, so that $\boldsymbol{Q}=Q\,{\boldsymbol{e}}_\perp$ (which is
sufficient for our purposes).%
\footnote{Previously ${\boldsymbol{e}}_\perp$ denoted a fixed arbitrary unit
$d-m$ vector. For convenience, we use here and below the same symbol
for the associated $d$ vector whose projection
onto the perpendicular subspace
yields the former while its $m$ parallel components vanish.}
For such values of  ${\boldsymbol{Q}}$,
the integral
$I_2({\boldsymbol{Q}})$ can be analytically calculated in a straightforward fashion,
either by going back to Eq.\ (16)  and (19) of I and computing the Fourier transform
of these distributions, or directly in momentum space, as we prefer to do here.
For dimensional reasons, we have
\begin{equation}\label{I2Q}
I_2(Q\,{\boldsymbol{e}}_{\perp})=Q^{-\epsilon}\,I_2({\boldsymbol{e}}_\perp)\;.
\end{equation}
The integral on the right-hand side is precisely the one written as  
$F_{m,\epsilon}/\epsilon$ in Eq.~(\ref{I2e}). Utilizing
a familiar method due to Feynman for folding two denominators into one (Eq.~(A8-1) of
Ref.~\cite{Ami84}), one is led to
\begin{eqnarray}\label{I2}
I_2({\boldsymbol{e}}_\perp)&=&{\int_0^1}\!\d{s}\int_{\boldsymbol{q}}
\left[
q_\|^4+q_\perp^2+2\,s\,{{\boldsymbol{q}}_\perp}\cdot{\boldsymbol{e}}_\perp
+s\,{\boldsymbol{e}}_\perp^2
\right]^{-2}\nonumber\\&=&
2^{-4+\frac{m}{2}+\epsilon}\,\pi^{-2+\frac{m}{4}+\frac{\epsilon}{2}}\,
\Gamma{\left(\frac{m+2\,\epsilon}{4}\right)}{\int_0^1\!}\d s
\int_{{\boldsymbol{q}}_\|}
{\left[q_\|^4+s(1-s)\right]}^{-{m+2\,\epsilon\over 4}}\nonumber\\&=&
(4\,\pi)^{-d/2}\,{\Gamma{\left(\frac{m}{4}\right)}\,\Gamma{\left(\frac{\epsilon}{2}\right)}\,
\Gamma^2{\left(1-\frac{\epsilon}{2}\right)}
\over 2\,\Gamma(2-\epsilon)\,\Gamma{\left(\frac{m}{2}\right)}}\;,
\end{eqnarray}
from which the result (\ref{Fmeps}) for $F_{m,\epsilon}$ follows at once.

The calculation of $I_4({\boldsymbol{Q}},{\boldsymbol{K}})$
is more involved; it is described in Appendix \ref{fourptgr},
giving
\begin{equation}\label{I4epsexp}
I_4(Q\,{\boldsymbol{e}}_\perp,{\boldsymbol{K}})
=F_{m,\epsilon}^2
\,{Q^{-2\epsilon}\over 2\epsilon}\,{\left[\frac{1}{\epsilon}
+J_u(m)+O(\epsilon)
\right]}\;,
\end{equation}
where $J_u(m)$ is the quantity defined in Eq.~(\ref{Ju}).

\subsection{Vertex function $\Gamma^{(2)}_{\phi^2}$}

Next, we turn to the vertex function $\Gamma^{(2)}_{\phi^2}$ with an insertion
of $\frac{1}{2}(\phi^2)_{{\boldsymbol{Q}}}=
\frac{1}{2}{\int}\d^dx\,\phi^2({\boldsymbol{x}})\,
\e^{i\,{\boldsymbol{Q}}\cdot{\boldsymbol{x}}}$.
To two-loop order it is given by%
\footnotemark[\value{footnote}]
\begin{eqnarray}
\Gamma^{(2)}_{\phi^2}({\boldsymbol{q}};\boldsymbol{Q})&=&1-\;
\raisebox{-3pt}{\begin{texdraw}\drawdim pt \setunitscale 2.5   \linewd 0.3
\lellip rx:2.8 ry:2.8
\move(0 2.8)\fcir f:0 r:0.5
\move(-3 -2.8) \rlvec(6 0)
\end{texdraw}}\;-\;
\raisebox{-3pt}{\begin{texdraw}\drawdim pt \setunitscale 2.5   \linewd 0.3
\lcir r:2.0 \move(0 4) \lcir r:2.0
\move(0 6)\fcir f:0 r:0.5
\move(-3 -2) \rlvec(6 0)
\end{texdraw}}\;-
\raisebox{-4.25pt}{
\begin{texdraw}
\drawdim pt \setunitscale 2.5   \linewd 0.3
\lellip rx:5 ry:2.8
\move(-8 0)
\lvec(8 0)
\move(0 2.8)
\fcir f:0 r:0.7
\end{texdraw}}
\;+O(u_0^3)\nonumber\\[1em]
&=&1-{\check{u}_0\over 2}\,I_2(\check{\boldsymbol{Q}})
+
{\check{u}^2_0}\,{\left[\frac{1}{4}\,I_2^2(\check{\boldsymbol{Q}})
+\frac{1}{2}\,
I_4(\check{\boldsymbol{Q}},\check{\boldsymbol{q}})\right]}+O(u_0^3)\;,
\end{eqnarray}
where the hatted momenta are again dimensionless ones, defined by analogy
with Eq.~(\ref{hatq}).

\section{Laurent expansion of the two-loop integral  
${\boldsymbol{I_4}}$}\label{fourptgr}

As can be seen from Eq.~(\ref{I4def}), the integral  
$I_4(\check{\boldsymbol{q}}_{12},\check{\boldsymbol{q}}_3)$
associated with the graph
\raisebox{-4pt}{\begin{texdraw}
\drawdim pt \setunitscale 1.5   \linewd 0.3
\move(-7 -1.5)\rlvec(15 6)
\move(-7 1.5)\rlvec(15 -6)
\move(5 0)
\lellip rx:1 ry:3
\end{texdraw}}
involves the divergent subintegral
$I_2(\check{\boldsymbol{q}}-\check{\boldsymbol{q}}_3)$.
The latter has a momentum-independent pole term
$\sim \epsilon^{-1}$ [cf.\
Eqs.~(\ref{I2}) and (\ref{Fmeps})].
Furthermore, the graph that results upon contraction
of this subgraph to a point [which
itself is proportional to $I_2(\check{\boldsymbol{q}}_{12}$)]
has contributions of order $\epsilon^0$ that
depend on $\check{q}_{12}$. Taken together, these
observations tell us that the pole term $\propto \epsilon^{-1}$
of $I_4(\check{\boldsymbol{q}}_{12},\check{\boldsymbol{q}}_3)$
depends on $\check{q}_{12}$ but not on $\check{q}_{3}$.%
\footnote{It is precisely this $\check{q}_{12}$-dependent
pole term that gets canceled by subtracting from the divergent
subgraph its pole part.}
Since the $\epsilon^{-2}$ pole of
$I_4(\check{\boldsymbol{q}}_{12},\check{\boldsymbol{q}}_3)$
is momentum independent, we can set $\check{q}_3=0$
when calculating the pole part of this integral.

To further simplify the calculation, we can choose
$\check{\boldsymbol{q}}_{12}$ to have
vanishing parallel component again, setting
$\check{\boldsymbol{q}}_{12}=Q\,{\boldsymbol{e}}_\perp$.
The integral to be calculated thus becomes
\begin{equation}\label{i2l}
I_4(Q\,{\boldsymbol{e}}_\perp;{\boldsymbol{0}})
={\int}\!\d^d{x}{\int}\!\d^d{y}\,  
G({\boldsymbol{y}})\,G({\boldsymbol{x}}-{\boldsymbol{y}})\,
\e^{i\,Q\,\boldsymbol{e}_\perp\cdot\boldsymbol{y}}
\,G^2({\boldsymbol{x}})\,,
\end{equation}
where $G({\boldsymbol{y}})$ now means the free propagator (\ref{Lprop})
with $\sigma_0=1$.
Let us substitute the free propagators
of the factor $G^2({\boldsymbol{x}})$
by their scaling form (\ref{Lsprop}) and
rewrite the Fourier integral ${\int}\!\d^d{y}\dots$ as a momentum-space
integral, employing the Schwinger representation (\ref{Schwinger}) for
both of the two free propagators in momentum space. Making
the change of variables
${\boldsymbol{x}}_\|\to{\boldsymbol{\upsilon}}={{\boldsymbol{x}}_\|}\,  
x_\perp^{-1/2}$, we arrive at
\begin{eqnarray}\label{start}
I_4(Q\,{\boldsymbol{e}}_\perp,{\boldsymbol{0}})
&=&{\int}\!\d^m \upsilon\, \Phi^2(\upsilon){\int}\!\d^{d-m}{x_\perp}\,  
x_\perp^{2\,\epsilon+{m\over 2}-4}\,
{\int_0^\infty}\!\d{s}{\int_0^\infty}\!\d{t}
\nonumber\\
&\times&
\int_{{\boldsymbol{q}}_\perp}\e^{-q_\perp^2\,(s+t) +
{{\boldsymbol{q}}_\perp}\cdot{(i\,\boldsymbol{x}_\perp-2\,  
t\,Q\,{\boldsymbol{e}}_\perp)}-t\,Q^2}\,
\int_{{\boldsymbol{q}}_\|}\e^{-q_\|^4\,(s+t)
+i\,{\boldsymbol{q}}_\|\cdot{\boldsymbol{\upsilon}}\sqrt{x_\perp}}.
\end{eqnarray}
Now the momentum integrations $\int_{{\boldsymbol{q}}_\perp}$
and $\int_{{\boldsymbol{q}}_\|}$ are decoupled and can be performed
in a straightforward fashion. That the latter integral takes such
a simple form is due to our choice of $\boldsymbol{Q}$ with
${\boldsymbol{Q}}_\|=\boldsymbol{0}$. Performing the angular integrations
yields
\begin{eqnarray}\label{Iqpar}
\lefteqn
{\int_{{\boldsymbol{q}}_\|}\e^{-q_\|^4\,(s+t)
+i\,{\boldsymbol{q}}_\|\cdot{\boldsymbol{\upsilon}}\sqrt{x_\perp}}}
&&\nonumber\\
&&=
(2\pi)^{-{m\over 2}}\int_0^\infty\!\d q_\|\,
q_\|^{m\over 2}\,\e^{-q_\|^4\,(s+t)}\, (\upsilon^2\,x_\perp)^{{2-m\over  
4}}J_{m-2\over 2}{\left(q_\| \,\upsilon\,x_\perp^{1/2}\right)}\;.
\end{eqnarray}
We replace the Bessel function in Eq.~(\ref{Iqpar}) by
its familiar Taylor expansion
\begin{equation}\label{bessel}
J_\mu(w)={\left({w\over 2}\right)}^\mu\,
\sum_{k=0}^\infty
{(-1)^k\,w^{2k}\over 2^{2k}\,{k!}\,\Gamma(\mu+k+1)}\,,
\end{equation}
integrate term by term over $q_\|$, employing
\begin{equation}
{\int_0^\infty}\!\d{q}\,q^{m -1+2\,k}\,\e^{-q^4\,(s+t)}=
{1 \over 4}\,\Gamma{\left({m+2\,k\over 4}\right)}\,\,(s+t)^{-{m+2\,k\over 4}}\,,
\end{equation}
and simplify the resulting ratio of $\Gamma$-functions by means of
the well-known duplication formula (6.1.18) of Ref.~\cite{AS72}.
This gives
\begin{equation}\label{Iqpar2}
{\int_{{\boldsymbol{q}}_\|}\e^{-q_\|^4\,(s+t)
+i\,{\boldsymbol{q}}_\|\cdot{\boldsymbol{\upsilon}}\sqrt{x_\perp}}}
=\pi^{{1-m\over 2}}\,
\sum_{k= 0}^\infty{{\left(-\upsilon^2 \, x_\perp\right)}^k\over {k!}\,  
\Gamma{\left({2+m+4k\over 4}\right)}}\,
{\left(8\,\sqrt{s+t}\right)}^{-{m+2\,k\over 2}}
\,.
\end{equation}
The integration over ${\boldsymbol{q}}_\perp$ in Eq.~(\ref{start}) is Gaussian.
Upon substituting the result together with the above equations
into (\ref{start}),
we get
\begin{eqnarray}\label{I4C}
I_4(Q\,{\boldsymbol{e}}_\perp,{\boldsymbol{0}})&=&
2^{\epsilon-4-m}\,\pi^{{2\epsilon-4-m\over 4}}
{\int}\!\d^m{\upsilon}\,
\Phi^2(\upsilon)\!\int_0^\infty\!\d{s}\!\int_0^\infty\!\d{t}\,
(s+t)^{{\epsilon-4\over 2}}\,\e^{-{s\,t\over s+t}\,Q^2}\nonumber\\[0.5em]
&&\times
\sum_{k=0}^\infty{C_k(Q;s,t)\over k!\, \Gamma{\left({2+m+2k\over 4}\right)}}\,
\left({-\upsilon^2\over 8\,\sqrt{s+t}} \right)^k
\end{eqnarray}
with
\begin{eqnarray}
C_k(Q;s,t)&\equiv&{\int}\!\d^{d-m} x_\perp\,
x_\perp^{k-4+2\,\epsilon+{m\over 2}}\,
{\exp}{\left[-{\,x_\perp^2+4\,i\,t\,Q\,
{{\boldsymbol{e}}_\perp}\cdot{\boldsymbol{x}}_\perp
\over 4\,( s+t)}\right]}\,.
\end{eqnarray}
We first perform the angular integrations and subsequently the
radial integration of the latter integral, obtaining
\begin{eqnarray}\label{Ck}
C_k(Q;s,t)
&=&
(2\,\pi)^{\vartheta_m}\,{\int_0^\infty}\!\d{r}\, r^{k-2+{m+6\,\epsilon\over 4}}\,
\e^{-{r^2\over 4\,(s+t)}}\,\left( {t\,Q\over s+t}\right)^{1-\vartheta_m}\,
{J_{\vartheta_m{-}1}}{\left( {t\,Q\,r\over s+t}\right)}\nonumber\\[0.5em]
&=&{\pi}^{\vartheta_m}\,
{2^{k+\epsilon}\,\Gamma{\left({k+\epsilon\over 2}\right)}\over \Gamma(\vartheta_m)}
\,(s+t)^{{k+\epsilon\over 2}}\, {_1\!F_1}{\left({k+\epsilon\over 2};
{\vartheta_m};{-t^2\,Q^2\over s+t} \right)}\,,
\end{eqnarray}
where we have introduced
\begin{equation}
\vartheta_m\equiv{d-m\over 2} = 2-{m\over 4}-{\epsilon\over 2}\,.
\end{equation}

Next we insert this result into expression (\ref{I4C}) for $I_4$,
and make a change of variable $s\to z=s/t$ The $t$-integration then
becomes straightforward (see Eq.~(2.22.3.1) of Ref.~\cite{PBM86}),
and we find that
\begin{eqnarray}\label{steps}
I_4(Q\,{\boldsymbol{e}}_\perp,{\boldsymbol{0}})&=&Q^{-2\,\epsilon}\,
{2^{2\,\epsilon-4-m}\,\pi^{{1-m\over 2}}\,\Gamma(\epsilon)
\over\Gamma{\left({8-m-2\,\epsilon\over 4}\right)}}
\nonumber\\&&
\times
{\int\!}\d^m{\upsilon}\, \Phi^2(\upsilon;m,d)\,
\sum_{k=0}^\infty A_k(m,\epsilon)\,(-\upsilon^2)^k\,,
\end{eqnarray}
with
\begin{eqnarray}\label{Ak}
A_k(m,\epsilon)&=&{\Gamma{\left({k+\epsilon\over 2}\right)}
\over k!\,2^{2k}\Gamma{\left({2+m+2\,k\over 4}\right)}}\,
{\int_0^\infty}\!\d{z}\,z^{-\epsilon}\,(z+1)^{2\epsilon-2}
\,{_{2\!}{F}_{1}}{\left(\epsilon\,,{k+\epsilon\over 2};\vartheta_m;{-1\over z}  
\right)}\,.\nonumber\\
\end{eqnarray}
Owing to the overall factor $\Gamma(\epsilon)$ and the additional
factor $\Gamma(\epsilon/2)$ of the coefficient $A_0(m,\epsilon)$,
the $k=0$ term of the above series contributes poles of second and first
order in $\epsilon$ to $I_4$. The remaining terms with $k\ge 1$ yield
poles of first order in $\epsilon$. Consider first the $k=0$ term.
The value of the  integral over $\upsilon$ may be gleaned from I
[cf.\ its Eqs.~(3.16) and (4.47)]:
\begin{equation}
{\int\!}\d^m{\upsilon}\, \Phi^2(\upsilon;m,d)=
{\frac{{2^{-5 - {m\over 2}}}\,{{\pi }^{\epsilon-4}}\,
     \Gamma{\left({\frac{m}{4}}\right)}\,
     \Gamma{\left(2 - {\frac{m}{4}} - \epsilon\right)}\,
     {{\Gamma^2{\left(1 - {\frac{\epsilon }{2}}\right)}}}}
     {\Gamma{\left({\frac{m}{2}}\right)}\,
     \Gamma(2 - \epsilon )}}\;.
\end{equation}
The integral over $z$ in $A_0(m,\epsilon)$ can be evaluated
explicitly by means of Mathematica \cite{MAT}. Alternatively, one
can change to the integration variable $\zeta=1/z$
and look up the transformed integral in Eq.~(2.21.1.15)
of the integral tables \cite{PBM86}. The result has a simple
expansion to order $\epsilon$, giving
\begin{equation}
A_0(m,\epsilon)={\Gamma{\left({\epsilon/ 2}\right)}
\over \Gamma{\left({2+m\over 4}\right)}}\,{\left[1+2\epsilon+O(\epsilon^2)\right]}
\end{equation}
upon substitution into Eq.~(\ref{Ak}).

Turning to the contributions with $k\ge 1$, we note that
both the scaling function
$\Phi(\upsilon;m,d)$ and the coefficients $A_k(m,\epsilon)$
may be taken at $d=d^*$ (i.e.\ $\epsilon=0$).
Then the integral ${\int_0^\infty}\!\d{z}\ldots$ reduces to one
and the series $\sum_{k=1}^\infty$ becomes the function
$\Theta(\upsilon;m)$ introduced in Eq.~(\ref{Theta}).
It follows that
\begin{equation}
{\int\!}\d^m{\upsilon}\, \Phi^2(\upsilon;m,d)\,\sum_{k=1}^\infty
A_k(m,\epsilon)\,(-\upsilon^2)^k={j_u(m)\over B_m}\,
{\left[1+O(\epsilon)\right]}\;,
\end{equation}
where $j_u(m)$ and $B_m$ are the integral (\ref{judef}) and
the coefficient (\ref{Bm}), respectively.
Combining the above results and expanding the prefactors of
the integral in Eq.~(\ref{steps}), we finally
obtain the result stated in Eq.~(\ref{I4epsexp}).

\section{Asymptotic behavior of $\Theta(\upsilon)$}\label{asbehTheta}

Upon differentiating the series (\ref{Theta}) of $\Theta(\upsilon;m)$
termwise and comparing with the Taylor expansion (\ref{sinsum})
of the scaling function $\hat{\Phi}$, one sees that the following relation holds:
\begin{equation}\label{Thetaprime}
\frac{\partial\Theta(\upsilon;m)}{\partial \upsilon}=\frac{4}{\upsilon}\,{\left[
\hat{\Phi}(\upsilon;m,d^*)-\hat{\Phi}(0;m,d^*)
\right]}\;.
\end{equation}
>From Eq.~(\ref{sinsum}) we can read off the value
$\hat{\Phi}(0;m,d^*)=1/\Gamma[(m+2)/ 4]$.
Let us substitute the asymptotic expansion Eq.~(\ref{asexpPhihat})
of $\hat{\Phi}(\upsilon;m,d^*)$ into this equation and integrate.
This yields
\begin{equation}
\Theta(\upsilon;m)\mathop{\approx}\limits_{\upsilon\to\infty}
{-4\,\ln\upsilon+C_\Theta(m)\over\Gamma{\left({m+2\over 4}\right)}}
+\sum_{k=1}^\infty
\frac{2\,(-1)^k\,\Gamma(2k)}
{k!\,\Gamma{\left(\frac{m+2 - 4k}{4}\right)}}
\,{\left({\upsilon\over 2}\right)}^{-4\,k}
\;.
\end{equation}
The terms of orders $\upsilon^{-4}$ and $\upsilon^{-8}$ agree with those
of the asymptotic form (\ref{eq:asbehTheta}) of $\Theta(\upsilon;m)$.
Hence it remains to show that
the integration constant $C_\Theta$ is given by
\begin{equation}\label{Cthetaval}
C_\Theta(m)=\psi{\left({m+2\over 4}\right)}
-C_E+\ln 16\;.
\end{equation}
To this end an integral representation of $\Theta(\upsilon;m)$ is
helpful. Consider the integral
\begin{equation}
J_\Theta(\upsilon;m,\epsilon)\equiv 2^{4+m}\,\pi^{{6+m-2\epsilon\over 4}}\,
{ \int_{{\boldsymbol{q}}}}
{\e^{i\,{{\boldsymbol{q}}_\|}\cdot{\boldsymbol{\upsilon}}+
i\,{{\boldsymbol{q}}_\perp}\cdot{\boldsymbol{e}}_\perp}
\over
{\big(q_\|^4+q_\perp^2\big)}^{2}}
=
\sum_{k=0}^\infty\frac{\Gamma{\left({k-\epsilon\over 2}\right)}\,(-\upsilon^2)^k}
{k!\,2^{2k}\,\Gamma{\left({2+m+2k\over 4}\right)}}\;,
\end{equation}
in terms of which $\Theta(\upsilon;m)$ can be written as
\begin{equation}\label{ThetaJTheta}
\Theta(\upsilon;m)=\lim_{\epsilon\to 0}\,{\left[J_\Theta(\upsilon;m,\epsilon)
-{\Gamma{\left(-{\epsilon\over 2}\right)}\over \Gamma{\left({2+m\over  
4}\right)}}\right]}
\end{equation}
and whose large-$\upsilon$ form
\begin{eqnarray}\label{JThetaas}
J_\Theta(\upsilon;m,\epsilon)
&\mathop{\approx}\limits_{\upsilon\to\infty}&
2^{4+m}\,\pi^{{6+m-2\epsilon\over 4}}\,\upsilon^{2\epsilon}
{ \int_{{\boldsymbol{q}}}}
{\e^{i\,{{\boldsymbol{q}}_\|}\cdot{\boldsymbol{e}}_\|}
\over
{\big(q_\|^4+q_\perp^2\big)}^{2}}\,{\left[1+O(\upsilon^{-4})\right]}
\nonumber\\
&=&\frac{2^{1-2\epsilon}\,\Gamma(-\epsilon)}{\Gamma{\left({2+m\over 4}\right)}}
\,\upsilon^{2\epsilon}\,{\left[1+O(\upsilon^{-4})\right]}
\end{eqnarray}
is easily derived. Insertion of the latter result into
Eq.~(\ref{ThetaJTheta}) gives the value (\ref{Cthetaval}) of $C_\theta$.

\section{Numerical integration}\label{numint}

The quantities $j_\phi(m)$, $j_\sigma(m)$, $j_\rho(m)$,
and $j_u(m)$ in terms of
which we  expressed the series expansion coefficients of the renormalization factors
and the critical exponents are integrals of the form
${\int_0^\infty}\!\d\upsilon\,f(\upsilon)$ [cf.\  
Eqs.~(\ref{jphidef})--(\ref{jrhodef}) and (\ref{judef})]. Their integrands,  
$f$, while integrable and decaying to zero
as $\upsilon\to\infty$, in general involve differences of generalized hypergeometric
functions, i.e., differences of  functions that grow exponentially as  
$\upsilon\to\infty$.
Therefore standard numerical integration procedures
run into problems when the upper integration limit becomes large.

To overcome this difficulty, we proceed in a similar manner as in I.
>From our knowledge of the asymptotic expansions of the
 functions $\Phi(\upsilon;m,d^*)$, $\Xi(\upsilon;m,d^*)$,
and $\Theta(\upsilon;m)$ we can determine that of the integrand.
Let $f^{(M)}_{\mathrm{as}}(\upsilon)$ be the asymptotic expansion
of $f(\upsilon)$ to order $\upsilon^{-M}$.
Then we have
\begin{equation}\label{asexpf}
f(\upsilon)-f^{(M)}_{\mathrm{as}}(\upsilon)
\mathop{\approx}\limits_{\upsilon\to\infty}
{\sum_{k=M+1}^\infty}C_f^{(k)}\,\upsilon^{-k}\;.
\end{equation}
We split the integrand as
\begin{equation}\label{Ifsplit}
{\int_0^\infty}\!{\d\upsilon}\,f(\upsilon)={\int_0^{\upsilon_0}}
\d\upsilon\,f(\upsilon)-
{\int^{\upsilon_0}_\infty}\!{\d\upsilon}\,f^{(M)}_{\mathrm{as}}(\upsilon)
+R_f^{(M)}(\upsilon_0)\;,
\end{equation}
where
\begin{equation}
R_f^{(M)}(\upsilon_0)\equiv {\int^\infty_{\upsilon_0}}\!{\d  
\upsilon}\,{\left[f(\upsilon)
-f^{(M)}_{\mathrm{as}}(\upsilon)\right]}\;.
\end{equation}
Then we choose $\upsilon_0$ as large as possible, but small enough so that
Mathematica \cite{MAT}  is still able to evaluate the integral
$\int_0^{\upsilon_0}\!f(\upsilon)\d\upsilon$ by numerical integration,
determine the second term on the right-hand side of Eq.~(\ref{Ifsplit})
by analytical integration, and neglect the third one. The
asymptotic expansion of the latter is easily deduced from Eq.~(\ref{asexpf}).
It reads
\begin{equation}
R_f^{(M)}(\upsilon_0)\,
\mathop{\approx}\limits_{\upsilon\to\infty}\,
{\sum_{k=M}^\infty}{C_f^{(k)}\over k}\,\upsilon_0^{-k}\;.
\end{equation}

Since the expansion (\ref{asexpf}) is only asymptotic, the value of $M$
must not be chosen too large. In practice, we utilized the asymptotic
expansions of $\Phi$, $\Xi$, and $\Theta$ up to the orders $\upsilon^{-12}$,
$\upsilon^{-10}$, and $\upsilon^{-8}$ explicitly shown in the
respective Eqs.~(\ref{asPhid*}), (\ref{asexpXid*}), and (\ref{eq:asbehTheta}),
and then truncated the resulting expression of the integrand
$f(\upsilon)$ consistently at the largest possible order. As upper integration
limit $\upsilon_0$ of the numerical integration we chose values between
$9$ and $10$.

As a consequence of the fact that all integrands $f(\upsilon)$ have
an explicit factor of $\upsilon^m$, the precision of our results
decreases as $m$ increases. Furthermore, the accuracy is greatest for
$j_\phi(m)$, whose integrand's asymptotic expansion starts with
$\upsilon^{-(13-m)}$, a particularly high power of $\upsilon^{-1}$.
The precision is lower for $j_\sigma(m)$ and  $j_u(m)$
 because their integrands involve either
four more powers of $\upsilon$ than that of $j_\phi$ or else
the function $\Theta(\upsilon)$ as a factor, whose asymptotic expansion starts with
a term $\sim \ln\upsilon$.

As a test of our procedure we can compare the numerical values of
the integrals it produces for $m=2$ and $m=6$
with the analytically known exact
results (\ref{jphi26})--(\ref{ju26}). The agreement
one finds is very impressive: Nine decimal digits of the exact results are  
reproduced (even for $m=6$) when the numerical integration is done
by means of the Mathematica\cite{MAT} routine `Nintegrate' with the option  
`WorkingPrecision=40'. However, we must not forget that the cases $m=2$ and
$m=6$ are special in that the asymptotic expansions of the functions
$\Phi$, $\Xi$, and $\Theta$---and hence those of the integrands---vanish
or truncate after the first term. Hence it would be too optimistic to
expect such extremely accurate results for other values of $m$.
In the worst cases (e.g., that of $j_\sigma(7)$ and $j_u(7)$), the fourth
decimal digit typically changes if $\upsilon$ is varied in the
range $9\ldots 11$. Therefore we are confident that
the first two decimal digits of the
$m=7$ values given in Table \ref{tab:jvalues}
are correct. For smaller values of $m$ the precision is greater.%
\footnote{For example, in the cases of $j_\sigma(4)$, $j_u(4)$, and
$j_\phi(4)$ changes
of $\upsilon_0\in [9,11]$ affect only
the respective last
decimal digits (in parentheses)
of the numerical values $j_\sigma(4)\simeq 20.0677(4)$,
$j_u(4)\simeq 0.80378728(5)$, and $j_\phi(4)\simeq 0.80378728(5)$.}

\end{document}